\documentclass[twocolumn,showpacs,preprintnumbers,amsmath,amssymb,pra,aps,superscriptaddress]{revtex4-1}

\usepackage{graphicx}
\usepackage{multirow}
\usepackage{amsmath}
\usepackage{amssymb}
\usepackage{soul}
\usepackage{verbatim}
\usepackage{bm}
\usepackage[colorlinks=true, linkcolor=blue, citecolor=gray]{hyperref}
\usepackage[table]{xcolor}
\usepackage{xspace}

\newcommand{\ket}[1]{\ensuremath{|#1\rangle}\xspace}
\newcommand{\bra}[1]{\ensuremath{\langle #1|}\xspace}

\newcommand{\bfr}{\mathbf{r}}
\newcommand{\bfR}{\mathbf{R}}

\newcommand{\ddroit}{{\rm d}}

\newcommand{\be}{\begin{eqnarray}}

\newcommand{\chatdag}{\hat{c}^{\dag}}  
\newcommand{\chat}{\hat{c}}  
\newcommand{\ham}{\hat{H}}  

\newcommand{\nsys}{N_{\mathrm{sys}}}  
\newcommand{\nmeas}{n_{\mathrm{m}}}  
\newcommand{\Nmeas}{N_{\mathrm{m}}}  
\newcommand{\nham}{N_{\mathrm{H}}}  
\newcommand{\nex}{N_{\mathrm{E}}}  
\newcommand{\numunitaries}{N_{\mathrm{U}}}  
\newcommand{\numterms}{N_{\mathrm{T}}}  
\newcommand{\timeunitary}{T_{\mathrm{U}}} 
\newcommand{\timeprep}{T_{\mathrm{P}}} 
\newcommand{\fillingfactor}{\eta}  
\newcommand{\Ktot}{K} 

\newcommand{\PP}{\mathbb{P}}

\newcommand{\Hh}{\mathcal{H}}
\newcommand{\Aaa}{\mathcal{A}}

\makeatletter
\newcommand\footnoteref[1]{\protected@xdef\@thefnmark{\ref{#1}}\@footnotemark}
\makeatother

\begin{document}

\title{Calculating energy derivatives for quantum chemistry on a quantum computer}

\author{T.E.~O'Brien}
\affiliation{Instituut-Lorentz, Universiteit Leiden, P.O. Box 9506, 2300 RA Leiden, The Netherlands}
\author{B.~Senjean}
\affiliation{Instituut-Lorentz, Universiteit Leiden, P.O. Box 9506, 2300 RA Leiden, The Netherlands}
\affiliation{Division of Theoretical Chemistry, Vrije Universiteit Amsterdam, De Boelelaan 1083, 1081 HV Amsterdam, The Netherlands}
\author{R.~Sagastizabal}
\affiliation{QuTech, Delft University of Technology, P.O. Box 5046, 2600 GA Delft, The Netherlands}
\affiliation{Kavli Institute of Nanoscience, Delft University of Technology, P.O. Box 5046, 2600 GA Delft, The Netherlands}
\author{X.~Bonet-Monroig}
\affiliation{Instituut-Lorentz, Universiteit Leiden, P.O. Box 9506, 2300 RA Leiden, The Netherlands}
\affiliation{QuTech, Delft University of Technology, P.O. Box 5046, 2600 GA Delft, The Netherlands}
\author{A.~Dutkiewicz}
\affiliation{Instituut-Lorentz, Universiteit Leiden, P.O. Box 9506, 2300 RA Leiden, The Netherlands}
\affiliation{Faculty of Physics, University of Warsaw, ul. Pasteura 5, 02–093 Warszawa, Poland}
\author{F.~Buda}
\affiliation{Leiden Institute of Chemistry, Leiden University, Einsteinweg 55, P.O. Box 9502, 2300 RA Leiden, The Netherlands}
\author{L.~DiCarlo}
\affiliation{QuTech, Delft University of Technology, P.O. Box 5046, 2600 GA Delft, The Netherlands}
\affiliation{Kavli Institute of Nanoscience, Delft University of Technology, P.O. Box 5046, 2600 GA Delft, The Netherlands}
\author{L.~Visscher}
\affiliation{Division of Theoretical Chemistry, Vrije Universiteit Amsterdam, De Boelelaan 1083, 1081 HV Amsterdam, The Netherlands}

\begin{abstract}
Modeling chemical reactions and complicated molecular systems has been proposed as the `killer application' of a future quantum computer.
Accurate calculations of derivatives of molecular eigenenergies are essential towards this end, allowing for geometry optimization, transition state searches, predictions of the response to an applied electric or magnetic field, and molecular dynamics simulations.
In this work, we survey methods to calculate energy derivatives, and present two new methods: one based on quantum phase estimation, the other on a low-order response approximation.
We calculate asymptotic error bounds and approximate computational scalings for the methods presented.
Implementing these methods, we perform the world's first geometry optimization on an experimental quantum processor, estimating the equilibrium bond length of the dihydrogen molecule to within $0.014$~Angstrom of the full configuration interaction value.
Within the same experiment, we estimate the polarizability of the H$_2$ molecule, finding agreement at the equilibrium bond length to within $0.06$~a.u. ($2\%$ relative error).
\end{abstract}
\maketitle

\section{Introduction}
Quantum computers are at the verge of providing solutions for certain classes of problems that are intractable on a classical computer~\cite{Preskill2018NISQ}. 
As this threshold nears, an important next step is to investigate how these new possibilities can be translated into useful algorithms for specific scientific domains.
Quantum chemistry has been identified as a key area where quantum computers can stop being science and start doing science~\cite{lloyd1996universal,reiher2017elucidating,
cao2018quantum,mcardle2018quantum}.
This observation has lead to an intense scientific effort towards developing and improving quantum algorithms for simulating time evolution~\cite{abrams1997simulation,
zalka1998simulating} and calculating ground state energies~\cite{aspuru2005simulated,kitaev1995quantum,
peruzzo2014variational,mcclean2016theory} of molecular systems.
Small prototypes of these algorithms have been implemented experimentally with much success~\cite{peruzzo2014variational,omalley2016scalable,kandala2017hardware,
santagati2018witnessing,colless2018computation}.
However, advances over the last century in classical computational chemistry methods, such as density functional theory (DFT)~\cite{dreizler1990density},
coupled cluster (CC) theory~\cite{shavitt2009many}, 
and quantum Monte-Carlo methods~\cite{booth2009fermion}, set a high bar for quantum computers to make impact in the field.

The ground and/or excited state energy is only one of the targets for quantum chemistry calculations. 
For many applications one also needs to be able to calculate the derivatives of the molecular electronic energy with respect to a change in the Hamiltonian~\cite{jensen2007introduction, norman2018principles}.
For example, the energy gradient (or first-order derivative) for nuclear displacements is used to search for minima, transition states, and reaction paths~\cite{schlegel2011geometry} that characterize a molecular potential energy surface (PES). They also form the basis for molecular dynamics (MD) simulations 
to dynamically explore the phase space of the system in its electronic ground state~\cite{marx2009AIMD} or, after a photochemical transition, in its electronically excited state~\cite{tully1990molecular}. While classical MD usually relies on force-fields which are parameterized on experimental data, there is a growing need to obtain these parameters on the basis of accurate quantum chemical calculations. One can easily foresee a powerful combination of highly accurate forces generated on a quantum computer with machine learning algorithms for the generation of reliable and broadly applicable force-fields~\cite{behler2016perspective}. This route might be particularly important in exploring excited state PES and non-adiabatic coupling terms, which are relevant in describing light-induced chemical reactions~\cite{li2014first,curchod2013trajectory,faraji2018calculations}. Apart from these perturbations arising from changing the nuclear positions, it is also of
interest to consider the effect that small external electric and/or magnetic fields have on the molecular energy. These determine well-known molecular properties, such as the (hyper)polarizability, magnetizability, A- and g-tensors, nuclear magnetic shieldings, among others.

Although quantum algorithms have been suggested to calculate derivatives of a function represented on a quantum register~\cite{jordan2005fast,gilyen2017optimizing,dallaire2018low,
schuld2018evaluating,harrow2019low}, or of derivatives of a variational quantum eigensolver (VQE) for optimization purposes~\cite{romero2018strategies,guerreschi2017practical}, 
the extraction of molecular properties from quantum simulation has received relatively little focus.
To the best of our knowledge only three investigations; in geometry optimization and molecular energy derivatives~\cite{kassal2009quantum}, molecular vibrations~\cite{mcardle2018vibration}, and the linear response function~\cite{roggero2018linear}; have been performed to date.

In this work, we survey methods for the calculation of molecular energy derivatives on a quantum computer.
We calculate estimation errors and asymptotic convergence rates of these methods, and detail the classical pre-and post-processing required to convert quantum computing output to the desired quantities.
As part of this, we detail two new methods for such derivative calculations.
The first involves simultaneous quantum phase and transition amplitude (or propagator) estimation, which we name 'propagator and phase estimation' (PPE).
The second is based on truncating the Hilbert space to an approximate (relevant) set of eigenstates, which we name the 'eigenstate truncation approximation' (ETA).
We use these methods to perform geometry optimization of the H$_2$ molecule on a superconducting quantum processor, as well as its response to a small electric field (polarizability), and find excellent agreement with the full configuration interaction (FCI) solution.

\section{Main}

Let $\ham$ be a Hamiltonian on a $2^{\nsys}$-dimensional Hilbert space (e.g.~the Fock space of an $\nsys$-spin orbital system), which has eigenstates
\begin{equation}
\ham|\Psi_j\rangle=E_j|\Psi_j\rangle,
\end{equation}
ordered by the energies $E_j$.
In this definition, the Hamiltonian is parametrized by the specific basis set that is used and has additional 
coefficients $\lambda_1,\lambda_2,\ldots$, which reflect fixed external influences on the electronic energy 
(e.g.~change in the structure of the molecule, or an applied magnetic or electric field).
An $d$th-order derivative of the ground state energy with respect to the parameters $\lambda_i$ is then defined as:
\begin{equation}
D_{\lambda_1,\lambda_2,\ldots}^{d_1,d_2,\ldots}=\frac{\partial^d E_0(\lambda_1,\lambda_2,\ldots)}{\partial^{d_1} \lambda_{1},\partial^{d_2} \lambda_{2},\ldots},\label{eq:deriv_def}
\end{equation}
where $d=\sum_i d_i$.
As quantum computers promise exponential advantages in calculating the ground state $E_0$ itself, it is a natural question to ask how to efficiently calculate such derivatives on a quantum computer.

\subsection{The quantum chemical Hamiltonian}
A major subfield of computational chemistry concerns solving the electronic structure problem.
Here, the system takes a second-quantized {\it ab initio} Hamiltonian, written in a basis of molecular spinors $\lbrace \phi_p(\bfr) \rbrace$ as follows:
\begin{eqnarray}\label{eq:fermionic_Hamil}
\ham = \sum_{pq} h_{pq} \hat{E}_{pq} + \dfrac{1}{2} \sum_{pqrs} g_{pqrs} \left( \hat{E}_{pq} \hat{E}_{rs} - \delta_{q,r} \hat{E}_{ps} \right),
\end{eqnarray}
where
$\hat{E}_{pq} =  \hat{c}_{p}^\dagger \hat{c}_{q}$
and $\hat{c}^{\dag}_{p}$ ($\hat{c}_{p}$) creates (annihilates) an electron in the molecular spinor $\phi_p$.
With equation (\ref{eq:fermionic_Hamil}) relativistic and non-relativistic realizations of the method only differ in the definition of the
matrix elements $h_{pq}$ and $g_{pqrs}$~\cite{visscher2002dirac}.
A common technique is to assume pure spin-orbitals and integrate over the spin variable. As we want to develop a formalism
that is also valid for relativistic calculations, we will remain working with spinors in this work. Adaptation to a spinfree formalism is straightforward, and will not affect computational scaling and error estimates.

The electronic Hamiltonian defined above depends parametrically on the nuclear positions, both explicitly via the nuclear potential and implicitly via the molecular orbitals that change when the nuclei are displaced.

\setlength{\tabcolsep}{6pt} 
\renewcommand{\arraystretch}{1.2} 

\begin{table*}
\begin{tabular}{|l|lr|lr|lr|lr|}\hline
 & \multicolumn{2}{c|}{Hellmann--Feynman} & \multicolumn{2}{c|}{PPE} & \multicolumn{2}{c|}{ETA} & \multicolumn{2}{c|}{Direct} \\
 & \multicolumn{2}{c|}{(first order)} & & & & & & \\\hline

 Time scaling &
    \cellcolor{yellow!50} $\epsilon^{-2}$&\cellcolor{yellow!50} $\dag$ &
    $\epsilon^{-2}$& * &
    \cellcolor{yellow!50} $\epsilon^{-2}$&\cellcolor{yellow!50} $\dag$ &
    $\epsilon^{-d+1}$& $\dag$ \\
with $\epsilon$ &
    $\epsilon^{-2}$& * &
    & &
    $\epsilon^{-2}$& * &
    $\epsilon^{-\frac{d+2}{2}}$& * \\
(fixed $\nsys$) &
    $\epsilon^{-1}$& ** &
    & &
    $\epsilon^{-1}$& ** &
    $\epsilon^{-1}$& $\triangledown$ \\\hline

 Time scaling &
    \cellcolor{yellow!50}$\nsys^4-\nsys^{13}$ &\cellcolor{yellow!50} $\dag$ &
    $\nsys^4-\nsys^{17.5}$ & * &
    \cellcolor{yellow!50}$\nsys^8-\nsys^{21}$ & \cellcolor{yellow!50} $\dag$ &
    $\nsys^4-\nsys^{13}$ & $\dag$\\
  with $\nsys$ &
    $\nsys^4-\nsys^{15}$ & *&
    & &
    $\nsys^8-\nsys^{23}$ & *&
    $\nsys^2-\nsys^7$ & *\\
 (approx, fixed $\epsilon$)&
    $\nsys^{3.5}-\nsys^{11}$ & ** & 
    & &
    $\nsys^{6.5}-\nsys^{17}$ & ** &
    $\nsys^2-\nsys^7$ & $\triangledown$ \\\hline
 Error sources & \multicolumn{2}{l|}{Basis set error} & \multicolumn{2}{l|}{Basis set error} & \multicolumn{2}{l|}{Basis set error} & \multicolumn{2}{l|}{Basis set error} \\
 unaccounted for & State error & ($\dag$,**) & \multicolumn{2}{l|}{Resolution error} & \multicolumn{2}{l|}{Truncation error} & State error & ($\dag$,**) \\\hline
Req. $\ham$ sim. & \multicolumn{2}{c|}{$*$, $**$} & \multicolumn{2}{c|}{All} & \multicolumn{2}{c|}{$*$, $**$} & \multicolumn{2}{c|}{$*$,$\triangledown$} \\\hline
\end{tabular}
\caption{\label{tab:method_summary}
Calculated performance of the energy derivative estimation methods suggested in this work. The computation time scaling as a function of the estimation error $\epsilon$ is given, alongside an approximate range of scalings with respect to the system size $\nsys$. 
Calculation details, full approximation details, and intermediate steps are given in Sec.~\ref{Methods.BoundCalc}.
Highlighted scalings correspond to the methods used in experiment in Sec.~\ref{sec:first_order}.
$\dag$ denotes performance estimates when using a VQE for state preparation, $*$ denotes performance when using QPE for state preparation, $**$ denotes performance if the amplitude amplification technique of~\cite{knill2007optimal} is used (which requires a VQE for state preparation), and $\triangledown$ denotes performing differentiation on a quantum computer with the methods of~\cite{kassal2009quantum,jordan2005fast} (which requires the ability to call the Hamiltonian as a quantum oracle as a function of the system parameters).
Details of errors not accounted for are given in Sec.~\ref{Methods.classical_calc.1} (basis set error), Sec.~\ref{Methods.GSErrorSystematic} (state error), Sec.~\ref{Methods.ResErr} (resolution error), and Sec.~\ref{Methods.LR} (truncation error).
We further note whether methods require phase estimation (which requires long coherence times).}
\end{table*}

\subsection{Asymptotic convergence of energy derivative estimation methods}\label{Main.Scalings}

In this section, we present and compare various methods for calculating energy derivatives on a quantum computer.
In Tab.~\ref{tab:method_summary}, we estimate the computational complexity of all studied methods in terms of the system size $\nsys$ and the estimation error $\epsilon$.
We also indicate which methods require quantum phase estimation, as these require longer coherence times than variational methods.
Many methods benefit from the amplitude estimation algorithm of~\cite{knill2007optimal}, which we have included costings for.
We approximate the scaling in $\nsys$ between a best-case scenario (a lattice model with a low-weight energy derivative and aggressive truncation of any approximations), and a worst-case scenario (the electronic structure problem with a high-weight energy derivative and less aggressive truncation).
The lower bounds obtained here are competitive with classical codes, suggesting that these methods will be tractable for use in large-scale quantum computing.
However, the upper bounds will need reduction in future work to be practical, e.g.~by implementing the strategies suggested in~\cite{mcclean2016theory,romero2018strategies,mcclean2017hybrid}.

For wavefunctions in which all parameters are variationally optimized, the Hellmann--Feynman theorem allows for ready calculation of energy gradients as the expectation value of the perturbing operator~\cite{kassal2009quantum,wecker2015solving}:
\begin{equation}
\frac{\partial E_0}{\partial\lambda}=\langle\Psi_0|\frac{\partial\ham}{\partial\lambda}|\Psi_0\rangle.\label{eq:FeynmanHellman}
\end{equation}
This expectation value may be estimated by repeated measurement of a prepared ground state on a quantum computer (Sec.~\ref{Methods.Prelim}), and classical calculation of the coefficients of the Hermitian operator $\partial\ham/\partial\lambda$ (Sec.~\ref{Methods.classical_calc.1}).
If state preparation is performed using a VQE, estimates of the expectation values in Eq.~\ref{eq:FeynmanHellman} will often have already been obtained during the variational optimization routine.
If one is preparing a state via QPE, one does not get these expectation values for free, and must repeatedly measure the quantum register on top of the phase estimation routine.
Such measurement is possible even with single-ancilla QPE methods which do not project the system register into the ground state (see Sec.~\ref{Methods.QPEbased.expval}).
Regardless of the state preparation method, the estimation error may be calculated by summing the sampling noise of all measured terms (assuming the basis set error and ground state approximation errors are minimal).

The Hellmann--Feynman theorem cannot be so simply extended to higher-order energy derivatives.
We now study three possible methods for such calculations.
The propagator and phase estimation (PPE) method uses repeated rounds of quantum phase estimation to measure the frequency-domain Green's function, building on previous work on Green's function techniques~\cite{dallaire2016method,bauer2016hybrid,roggero2018linear}.
We may write an energy derivative via perturbation theory as a sum of products of path amplitudes $A$ and energy coefficients $f_A$.
For example, a second order energy derivative may be written as
\begin{align}
&\frac{\partial^2 E_0}{\partial\lambda_1\partial\lambda_2}=\langle\Psi_0|\frac{\partial^2\ham}{\partial\lambda_1\partial\lambda_2}|\Psi_0\rangle\nonumber\\&+\sum_{j\neq 0}2\;\mathrm{Re}\left[\langle\Psi_0|\frac{\partial\ham}{\partial\lambda_1}|\Psi_j\rangle\langle\Psi_j|\frac{\partial\ham}{\partial\lambda_2}|\Psi_0\rangle\right]\frac{1}{E_0-E_j},\label{eq:second_order_deriv}
\end{align}
allowing us to identify two amplitudes
\begin{align}
\Aaa_1(j)&=\langle\Psi_0|\frac{\partial\ham}{\partial\lambda_1}|\Psi_j\rangle\langle\Psi_j|\frac{\partial{\ham}}{\partial\lambda_2}|\Psi_0\rangle,\\
\Aaa_2&=\langle\Psi_0|\frac{\partial^2\ham}{\partial\lambda_1\partial\lambda_2}|\Psi_0\rangle, \label{eq:A2}
\end{align}
and two corresponding energy coefficients
\begin{equation}
f_{1}(E_0;E_j)=\frac{2}{E_0-E_j},\hspace{0.2cm} f_2=1.
\end{equation}
The generic form of a $d$-th order energy derivative may be written as
\begin{align}
D=&\sum_{\Aaa}\sum_{j_1,\ldots,j_{X_{\Aaa}-1}}\;\mathrm{Re}\left[\Aaa(j_1,\ldots,j_{X_{\Aaa}-1})\right]\nonumber\\&\times f_{\Aaa}(E_0;E_{j_1},\ldots,E_{j_{X_{\Aaa}-1}}),\label{eq:Greens_functions}
\end{align}
where $X_{\Aaa}$ counts the number of excitations in the path.
As this is different from the number of responses of the wavefunction, $X_{\Aaa}$ does not follow the $2n+1$ rule; rather, $X_{\Aaa}\leq d$.
The amplitudes $\Aaa$ take the form~\footnote{Higher order ($d\geq 4$) amplitudes will eventually contain terms corresponding to disconnected excitations, which then are products of multiple terms of the form of Eq.~\ref{eq:Amplitudes}. Our procedure may be extended to include these contributions; we have excluded them here for the sake of readibility.}
\begin{align}
\Aaa(j_1&,\ldots,j_{X_{\Aaa}-1})\nonumber\\&=\langle\Psi_0|\hat{P}_{X_{\Aaa}}\prod_{x=1}^{X_{\Aaa}-1}\left(|\Psi_{j_x}\rangle\langle\Psi_{j_x}|\hat{P}_x\right)|\Psi_0\rangle.\label{eq:Amplitudes}
\end{align}
These may be estimated simultaneously with the corresponding energies $E_{j_x}$ by applying rounds of QPE in between excitations by operators $\hat{P}_x$ (Sec.~\ref{Methods.QPEGFCircuit}). One may then classically calculate the energy coefficients $f_A$, and evaluate Eq.~\ref{eq:Greens_functions}.
Performing such calculation over an exponentially large number of eigenstates $|\Psi_{j_x}\rangle$ would be prohibitive.
However, the quantum computer naturally bins small amplitudes of nearby energy with a resolution $\Delta$ controllable by the user.
We expect the resolution error to be smaller than the error in estimating the amplitudes $\Aaa(j_1,\ldots,j_{X_{\Aaa}-1})$ (Sec.~\ref{Methods.QPEbased.sampling}); we use the latter for the results in Tab.~\ref{tab:method_summary}.

\begin{figure}
\includegraphics[width=\columnwidth]{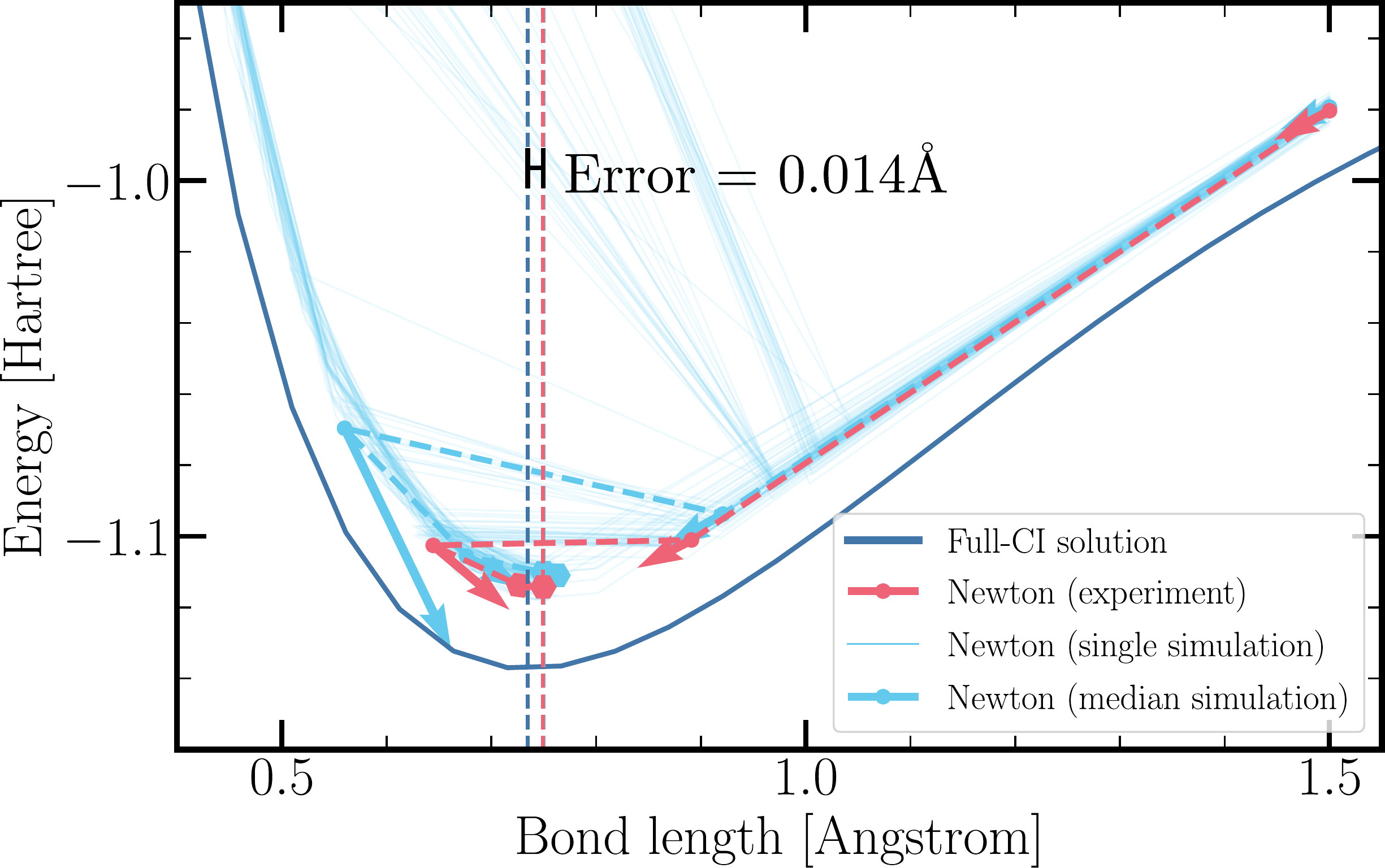}
\caption{\label{fig:single_gradient_descent}Illustration of geometry optimization of the H$_2$ molecule. A classical optimization algorithm (Newton) minimizes the estimation of the true ground state energy (dark blue curve) on a superconducting transmon quantum computer (red crosses) as a function of the bond distance $R_{\rm H-H}$. To improve convergence, the quantum computer provides estimates of the FCI gradient (red arrows) and the Hessian calculated with the response method.
Dashed vertical lines show the position of the FCI and estimated minima (error $0.014$\r{A}). Light blue dashed lines show the median value of $100$ density matrix simulations (Sec.~\ref{Methods.quantumsim}) of this optimization, with the shaded region the corresponding interquartile range.}
\end{figure}

In lieu of the ability to perform the long circuits required for phase estimation, one may approximate the sum over (exponentially many) eigenstates $|\Psi_j\rangle$ in Eq.~\ref{eq:Greens_functions} by taking a truncated set of (polynomially many) approximate eigenstates $|\tilde{\Psi}_j\rangle$.
We call such an approximation the eigenstate truncation approximation, or ETA for short.
However, on a quantum computer, we expect both to better approximate the true ground state $|\Psi_0\rangle$, and to have a wider range of approximate excited states~\cite{parrish2019quantum,mcclean2017hybrid,santagati2018witnessing,higgott2018variational,endo2018variational}.
In this work, we focus on the quantum subspace expansion (QSE) method of~\cite{mcclean2017hybrid}.
This method proceeds by generating a set of $\nex$ vectors $|\chi_j\rangle$ connected to the ground state $|\Psi_0\rangle$ by excitation operators $\hat{E}_j$,
\begin{equation}
|\chi_j\rangle=\hat{E}_j|\Psi_0\rangle.
\end{equation}
This is similar to truncating the Hilbert space using a linear excitation operator in the (classical) equation of motion coupled cluster (EOMCC) approach~\cite{Bartlett2011EOM}.
The $|\chi_j\rangle$ states are not guaranteed to be orthonormal; the overlap matrix
\begin{equation}
S^{(\mathrm{QSE})}_{j,k}=\langle\chi_j|\chi_k\rangle,\label{eq:S_red}
\end{equation}
is not necessarily the identity.
To generate the set $|\tilde{\Psi}_j\rangle$ of orthonormal approximate eigenstates, one can calculate the projected Hamiltonian matrix
\begin{equation}
H^{(\mathrm{QSE})}_{j,k}=\langle\chi_j|\ham|\chi_k\rangle,\label{eq:H_red}
\end{equation}
and solve the generalized eigenvalue problem:
\begin{equation}
\ham^{(\mathrm{QSE})}\vec{v}^{(j)}=\tilde{E}_j\hat{S}^{(\mathrm{QSE})}\vec{v}^{(j)}\rightarrow |\tilde{\Psi}_j\rangle=\sum_{l}\vec{v}^{(j)}_l|\chi_l\rangle.\label{eq:gen_eval}
\end{equation}

Regardless of the method used to generate the eigenstates $|\tilde{\Psi}_j\rangle$, the dominant computational cost of the ETA is the need to estimate $\nex^2$ matrix elements.
Furthermore, to combine all matrix elements with constant error requires the variance of each estimation to scale as $\nex^{-2}$ (assuming the error in each term is independent).
This implies that, in the absence of amplitude amplification, the computational complexity scales as $\nex^4$.
Taking all single-particle excitations sets $\nex\propto\nsys^2$.
However, in a lattice model one might consider taking only local excitations, setting $\nex\propto\nsys$.
Further reductions to $\nex$ will increase the systematic error from Hilbert space truncation (Sec.~\ref{Methods.LR}), although this may be circumvented somewhat by extrapolation.

For the sake of completeness, we also consider here the cost of numerically estimating an energy derivative by estimating the energy at multiple points;
\begin{align}
\frac{\partial^2E}{\partial\lambda^2}&=\frac{1}{\delta\lambda^2}(E(\lambda-\delta\lambda)+E(\lambda+\delta\lambda)-2E(\lambda))+O(\delta\lambda^2)\label{eq:second_order_finite_diff}\\
&=\frac{1}{\delta\lambda}\left(\frac{\partial E}{\partial\lambda}(\lambda+\delta\lambda/2)-\frac{\partial E}{\partial\lambda}(\lambda-\delta\lambda/2)\right)+O(\delta\lambda^2).
\end{align}
Here, the latter formula is preferable if one has direct access to the derivative in a VQE via the Hellmann--Feynman theorem, whilst the former is preferable when one may estimate the energy directly via QPE.
In either case, the sampling noise (Sec.~\ref{Methods.Prelim} and Sec.~\ref{Methods.QPEDef}) is amplified by the division of $\delta\lambda$.
This error then competes with the $O(\delta\lambda^2)$ finite difference error, the balancing of which leads to the scaling laws in Tab.~\ref{tab:method_summary}.
This competition can be negated by coherently copying the energies at different $\lambda$ to a quantum register of $L$ ancilla qubits and performing the finite difference calculation there~\cite{kassal2008polynomial,jordan2005fast}.
Efficient circuits (and lower bounds) for the complexity of such an algorithm have not been determined, and proposed methods involve coherent calculation of the Hamiltonian coefficients on a quantum register.
This would present a significant overhead on a near-term device, but with additive and better asymptotic scaling than the QPE step itself (which we use for the results in Tab.~\ref{tab:method_summary}).

\subsection{Geometry optimization on a superconducting quantum device}\label{sec:first_order}

To demonstrate the use of energy derivatives directly calculated from a quantum computing experiment, we first perform geometry optimization of the diatomic H$_2$ molecule, using two qubits of a superconducting transmon device.
(Details of the experiment are given in Sec.~\ref{Methods.expt}.)
Geometry optimization aims to find the ground state molecular geometry by minimizing the ground state energy $E_0(\bfR)$ as a function of the atomic co-ordinates $R_i$.
In this small system, rotational and translational symmetries reduce this to a minimization as a function of the bond distance $R_{\rm H-H}$
In Fig.~\ref{fig:single_gradient_descent}, we illustrate this process by sketching the path taken by Newton's minimization algorithm from a very distant initial bond distance ($R_{\rm H-H}=1.5$\r{A}).
At each step of the minimization we show the gradient estimated via the Hellman--Feynman theorem.
Newton's method additionally requires access to the Hessian, which we calculated via the ETA (details given in Sec.~\ref{Methods.expt}).
The optimization routine takes $5$ steps to converge to a minimum bond length of $0.749$\r{A}, within $0.014$\r{A} of the target FCI equilibrium bond length (given the chosen STO-3G basis set).
To demonstrate the optimization stability, we performed $100$ simulations of the geometry optimization experiment on the \emph{quantumsim} density-matrix simulator~\cite{obrien2017density}, with realistic sampling noise and coherence time fluctuations (details given in Sec.~\ref{Methods.quantumsim}).
We plot all simulated optimization trajectories on Fig.~\ref{fig:single_gradient_descent}, and highlight the median $(R_{\rm H-H},E(R_{\rm H-H}))$ of the first $7$ steps.
Despite the rather dramatic variations between different gradient descent simulations, we observe all converging to within similar error bars, showing that our methods are indeed stable.

\begin{figure}
\includegraphics[width=\columnwidth]{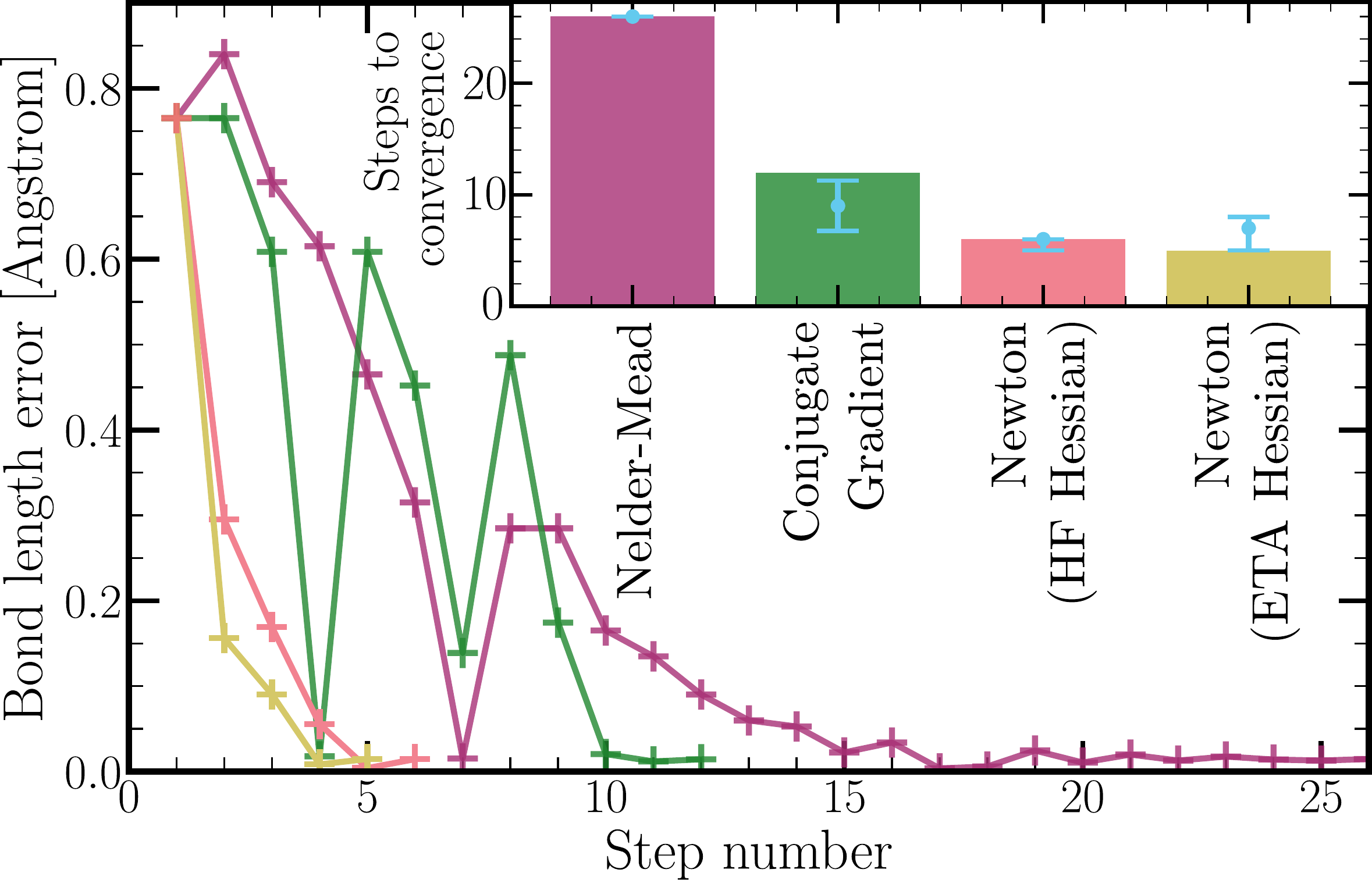}
\caption{\label{fig:gradient_descent_comparison}Comparison of geometry optimization via different classical optimization routines, using a quantum computer to return energies and Jacobians as required, and estimating Hessians as required either via the ETA on the experimental device, or the Hartree-Fock (HF) approximation on a classical computer. Each algorithm was run till termination with a tolerance of $10^{-3}$, so as to be comparable to the final error in the system. (Inset) bar plot of the number of function evaluations of the four compared methods. Light blue points correspond to median $N_{\mathrm{fev}}$ from $100$ density-matrix simulations (Sec.~\ref{Methods.quantumsim}) of geometry optimization, and error bars to the interquartile ranges.}
\end{figure}

To study the advantage in geometry optimization from direct estimation of derivatives on a quantum computer, we compare in Fig.~\ref{fig:gradient_descent_comparison} our performance with gradient-free (Nelder-Mead) and Hessian-free (conjugate gradient, or CG) optimization routines.
We also compare the performance of Newton's method with an approximate Hessian from Hartree-Fock (HF) theory.
All methods converge to near-identical minima, but both Newton methods converge about twice as fast as the raw CG method, which in turn converges about twice as fast as Nelder-Mead.
The density-matrix simulations predict that the ETA method Hessians provide less stable convergence than the HF Hessians; we attribute this to the fact that the HF Hessian at a fixed bond distance does not fluctuate between iterations.
The density-matrix simulations also predict the CG method performance to be on average much closer to the Newton's method performance.
However, we expect the separation between gradient and Hessian-free optimization routines to become more stark at larger system sizes, as is observed typically in numerical optimization~\cite{nocedal2006numerical}.

\begin{figure}
\includegraphics[width=\columnwidth]{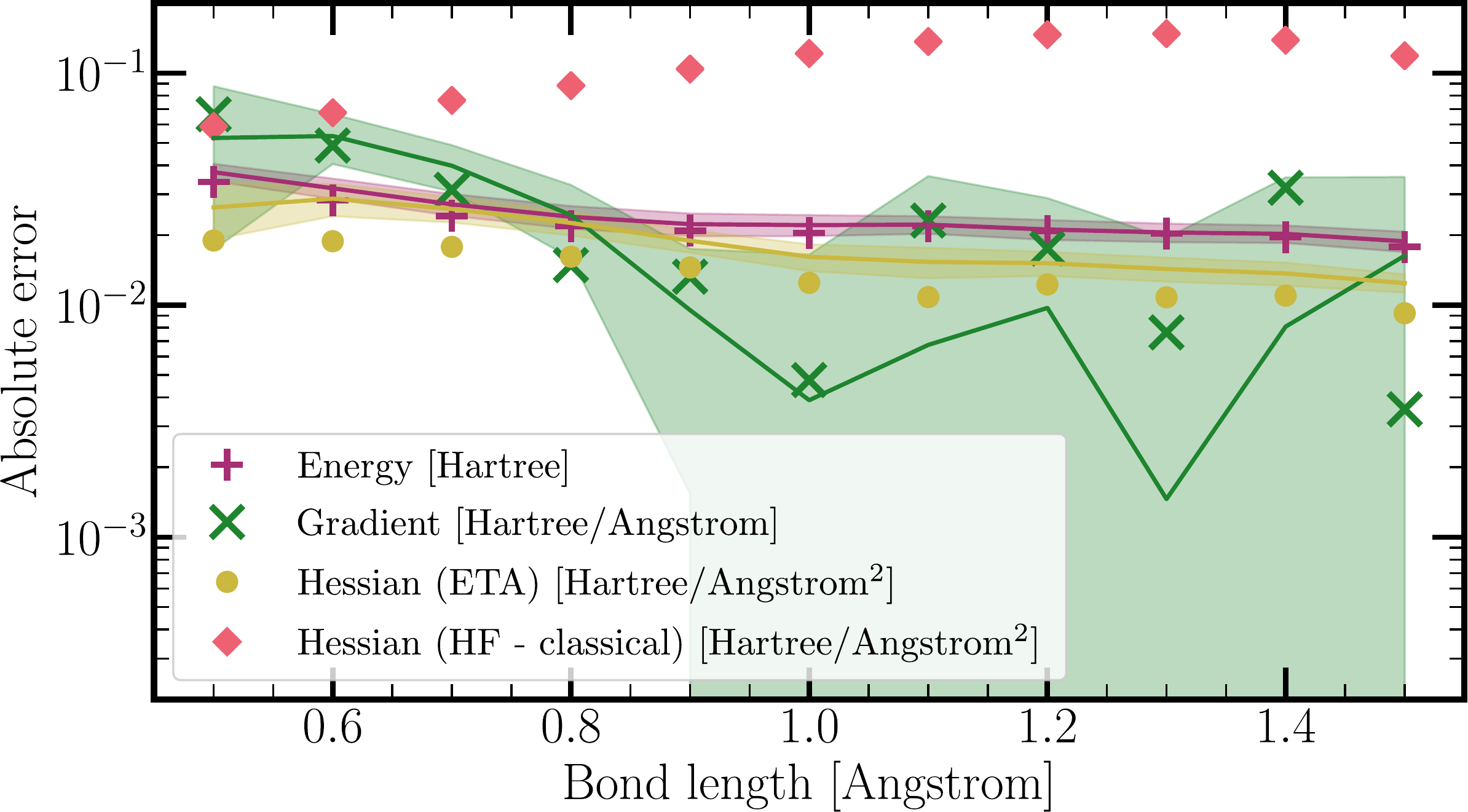}
\caption{\label{fig:abs_error}Absolute error in energies and energy derivatives from an experimental quantum computation on $11$ points of the bond dissociation curve of H$_2$.
The error is dominated here by experimental sources (in particular qubit decay channels); error bars from sampling noise are smaller than the points themselves.
Continuous lines connect the median value of $100$ density matrix simulations at each points, with the shaded region corresponding to errors to the interquartile range.}
\end{figure}

To separate the performance of the energy derivative estimation from the optimization routine, we study the error in the energy $E$, the Jacobian $J$ and Hessian $K$ given as
$\epsilon_A={|A_{\mathrm{FCI}}-A_{\mathrm{expt}}|}$, $(A=E,J,K).$
In Fig.~\ref{fig:abs_error}, we plot these errors for different bond distances.
For comparison we additionally plot the error in the HF Hessian approximation.
We observe that the ETA Hessian is significantly closer than the HF-approximated Hessian to the true value, despite the similar performance in geometry optimization.
The accuracy of the ETA improves at large bond distance, where the HF approximation begins to fail, giving hope that the ETA Hessian will remain appropriate in strongly correlated systems where this occurs as well.

\subsection{Polarizability estimation}

\begin{figure}
\includegraphics[width=\columnwidth]{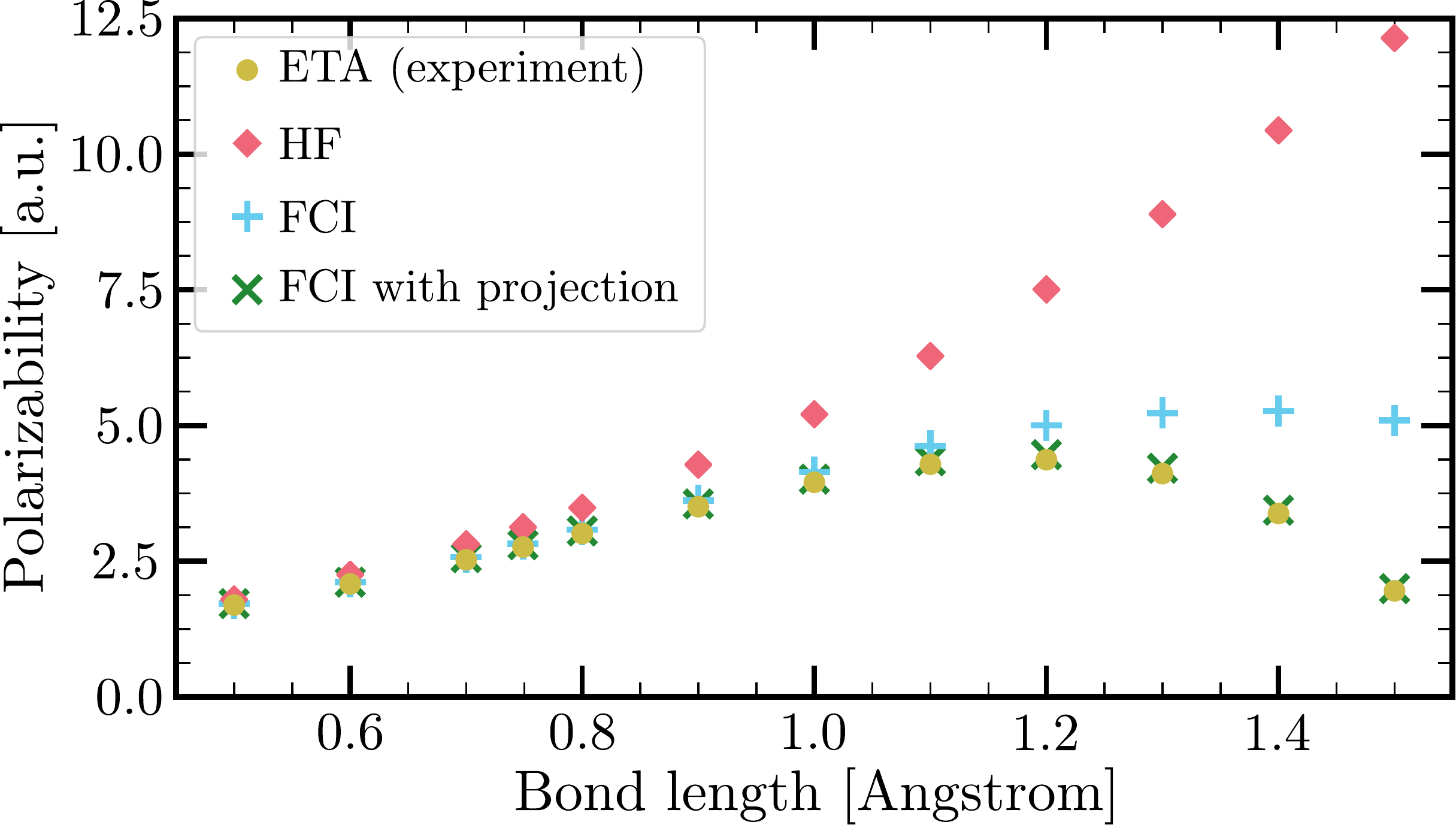}
\caption{\label{fig:polarizability}Estimated polarizability of the hydrogen molecule as a function of the bond distance, in atomic units (1 a.u. = 0.14818471~\r{A}$^3$).}
\end{figure}

A key property to model in quantum chemistry is the polarizability, which describes the tendency of an atom or molecule to acquire an induced dipole moment due to a change in an external electric field $\vec{F}$.
The polarizability tensor may be calculated as $\alpha_{i,j}=\left.\frac{\partial E(\vec{F})}{\partial F_i\partial F_j}\right|_{\vec{F}=0}$~\footnote{The first-order derivative $\partial E/\partial F_i$ gives the dipole moment, which is also of interest, but is zero for the hydrogen molecule.}.
In Fig.~\ref{fig:polarizability}, we calculate the $z$-component of the polarizability tensor of H$_2$ in the ETA, and compare it to FCI and HF polarizability calculations on a classical computer.
We observe good agreement to the target FCI result at low $R_{\rm H-H}$, finding a $0.060$ a.u. ($2.1\%$) error at the equilibrium bond distance (including the inaccuracy in estimating this distance).
However our predictions deviate from the exact result significantly at large bond distance ($R_{\rm H-H} \gtrsim 1.2$~\r{A}).
We attribute this deviation to the transformation used to reduce the description of H$_2$ to a two-qubit device (see Sec.~\ref{Methods.expt}), which is no longer valid when adding the dipole moment operator to the Hamiltonian.
To confirm this, we classically compute the FCI polarizability following the same transformation (which corresponds to projecting the larger operator onto a $2$-qubit Hilbert space).
We find excellent agreement between this and the result from the quantum device across the entire bond dissociation curve.
This implies that simulations of H$_2$ on a $4$-qubit device should match the FCI result within experimental error.

\section{Conclusion}
In this work, we have surveyed possible methods for estimating energy gradients on a quantum computer, including two new techniques of our own design.
We have estimated the computational complexity of these methods, both in terms of the accuracy required for the result and the size of the studied system.
We have demonstrated the use of these methods on a small-scale quantum computing experiment, obtaining the equilibrium bond length of the H$_2$ molecule to $0.014$\r{A} ($2\%$) of the target Full-CI value, and estimating the polarizability at this bond length to within $0.060$ a.u. ($2.1\%$) of the same target.

Our methods do not particularly target the ground state over any other eigenstate of the system, and so can be used out-of-the-box for gradient estimation for excited state chemistry.
They hold further potential for calculating frequency-domain Green's functions in strongly correlated physics systems (as PPE estimates the gradient through a Green's function calculation).
However, throughout this work we made the assumption that the gap $\delta$ between ground and excited state energies was sufficiently large to not be of concern (namely that $\delta\propto\nsys^{-1}$).
Many systems of interest (such as high-temperature superconductors) are characterized by gap closings in the continuum limit.
How this affects PPE is an interesting question for future work.
Further investigation is also required to improve some of the results drawn upon for this work, in particular reducing the number of measurements required during a VQE and improving amplitude estimation during single-round QPE.

\section{Methods}\label{Methods}

\subsection{Classical computation}\label{Methods.classical_calc}

The one- and two-electron integrals defining the fermionic
Hamiltonian in Eq.~\ref{eq:fermionic_Hamil}
are obtained from a preliminary HF calculation that is assumed to be easily feasible on a classical computer.
In non-relativistic theory the one-electron integrals are given by
\begin{eqnarray}\label{eq:1e_int}
h_{pq} = \int \ddroit \bfr \phi_p^* (\bfr)  \left( - \dfrac{1}{2} \nabla_{\bfr} + V(\bfr) \right) \phi_q(\bfr), \nonumber \\
\end{eqnarray}
where $V(\bfr)$ is the electron-nuclear attraction potential from fixed nuclei at positions $\bfR_i$.
The two-electron integrals are given by,
\begin{eqnarray}\label{eq:2e_int}
g_{pqrs} &=& \iint\ddroit \bfr_1 \ddroit \bfr_2 
\dfrac{\phi_p^*(\bfr_1)\phi_q(\bfr_1)\phi_r^*(\bfr_2)\phi_s(\bfr_2)}{r_{12}}.
\end{eqnarray}
For simplicity we used a finite  difference technique to compute the
matrix representations of perturbations corresponding to a change in 
nuclear coordinates and an external electric field
\begin{eqnarray}
\dfrac{\partial \ham}{\partial \lambda} \approx
\dfrac{\ham(\lambda + \delta\lambda/2)
- \ham(\lambda - \delta\lambda/2)}{\delta\lambda},
\end{eqnarray}
and
\begin{eqnarray}
\dfrac{\partial^2 \ham}{\partial \lambda^2} \approx
\dfrac{\ham(\lambda + \delta\lambda)
+ \ham(\lambda - \delta\lambda) - 2\ham(\lambda)}{\delta\lambda^2},
\end{eqnarray}
where $\delta\lambda = 0.001$ corresponds to a small change in $\lambda$.
The above (perturbed) quantum chemical Hamiltonians have been determined within the Dirac program~\cite{DIRAC18} and transformed into qubit Hamiltonians using the OpenFermion~\cite{mcclean2017openfermion} package.
This uses the newly-developed, freely-available~\cite{openfermion_dirac} OpenFermion-Dirac interface, allowing for the simulation of relativistic quantum chemistry calculations on a quantum computer.
While a finite difference technique was sufficient for the present purpose, such schemes are sensitive to numerical noise and 
have a high computational cost when applied to larger molecular systems. A consideration of the analytical calculation of energy derivatives 
can be found in the Supplementary Materials.

\subsection{Approximate bound calculation details}\label{Methods.BoundCalc}

In this section we detail our method for calculating the approximate bounds in Table.~\ref{tab:method_summary}.
We first estimate the error $\epsilon$ (Tab.~\ref{tab:method_summary_extd}, first row; details of the non-trivial calculations for the PPE and ETA methods given in Sec.~\ref{Methods.Second.QPE} and Sec.~\ref{Methods.LR}) respectively.
Separately, we may calculate the time cost by multiplying the number of circuits, the number of repetitions of said circuits ($\nmeas$, $\Nmeas$, and $\Ktot$ depending on the method), and the time cost of each circuit (Tab.~\ref{tab:method_summary_extd}, second row).
(This assumes access to only a single quantum processor, and can in some situations be improved by simultaneous measurement of commuting terms, as discussed in Sec.~\ref{Methods.Prelim}.)
We then choose the scaling of the number of circuit repetitions as a function of the other metaparameters to fix $\epsilon$ constant (Tab.~\ref{tab:method_summary_extd}, third row).
We finally substitute the lower and upper bounds for these metaparameters in terms of the system size as stated throughout the remaining sections.
For reference, we summarize these bounds in Tab.~\ref{tab:approx_summary}.

\begin{table*}
\begin{tabular}{|l|lr|lr|lr|lr|}\hline
 & \multicolumn{2}{c|}{Hellmann--Feynman} & \multicolumn{2}{c|}{PPE} & \multicolumn{2}{c|}{ETA} & \multicolumn{2}{c|}{Direct} \\
 & \multicolumn{2}{c|}{(first order)} & & & & & & \\\hline
 Error scaling ($\epsilon$) &
    $\numterms^{\frac{1}{2}}\nmeas^{-\frac{1}{2}}\;\;$ & $\dag$ &
    $\numunitaries^{\frac{d}{2}}\times$ & *&

    $\nex \numterms^{\frac{1}{2}}\nmeas^{-\frac{1}{2}}\;\;$ & $\dag$ &
    $(\numterms\nmeas^{-1})^{\frac{1}{d+1}}\;\;$ & $\dag$\\
 &
    $\numunitaries^{\frac{1}{2}}\Nmeas^{-\frac{1}{2}}A_0^{-\frac{1}{2}}$ & * &
    \multicolumn{2}{l|}{
    $\left(d^{\frac{d}{2}}\delta^{d-2}\Delta^{\frac{d}{2}}K^{-1}t^{-1}A_0^{-1}\right.$} &
    $\nex\numunitaries^{\frac{1}{2}}\Nmeas^{-\frac{1}{2}}A_0^{-\frac{1}{2}}$ & * &
    $(\Ktot A_0t)^{\frac{-2}{d+2}}$ & * \\
 &
    $\numunitaries^{\frac{1}{2}}\Ktot^{-1}$ & ** &
    \multicolumn{2}{l|}{
    $\left.\vphantom{\delta^{-1}\Ktot^{\frac{d}{4}-\frac{3}{2}}}+\delta^{d-1}\Delta^{\frac{d}{4}}\Nmeas^{-\frac{1}{2}}A_0^{-\frac{1}{2}}\right)$} &
    $\nex \numunitaries^{\frac{1}{2}}\Ktot^{-1}$ & ** &
    $(KA_0t)^{-1}$& $\triangledown$ \\\hline
 Time scaling & 
    $\numterms\nmeas\timeprep$ & $\dag$ & 
    $d\numunitaries^d\Ktot\timeunitary$ & * & 
    $\numterms\nex^2\nmeas\timeprep$ & $\dag$ &
    $d\numterms\nmeas\timeprep$ & $\dag$ \\
 &
    $\numunitaries\Ktot\timeunitary$ & * &
    & &
    $\numunitaries\nex^2\Ktot\timeunitary$ & * &
    $d\Ktot\timeunitary$ & * \\
&
    $\numunitaries\Ktot\timeprep$ & ** &
    & &
    $\numunitaries\nex^2\Ktot\timeprep$ & ** &
    $2^d KT_U$ & $\triangledown$ \\\hline
 Time scaling &
    $\numterms^2\timeprep$ & $\dag$ &
    $\times\timeunitary$ & * &

    $\numterms^2\nex^4\timeprep $ & $\dag$ & 
    $d\numterms^2\timeprep $ & $\dag$\\
at fixed $\epsilon$ &
    $A_0^{-2}N_U^2\timeunitary $ & *&
    \multicolumn{2}{l|}{$\left(d^{\frac{3}{2}}\delta^{d-2}t^{-1}\Delta^{\frac{d}{2}}A_0^{-1}\numunitaries^{\frac{3d}{2}-1}\right.$} & 
    $\numunitaries^2\nex^4A_0^{-2}\timeunitary$ & *&
    $d t^{-1}A_0^{-1}\timeunitary $ & *\\
 &
    $\numunitaries^{\frac{3}{2}}\timeprep$ & ** &
    \multicolumn{2}{l|}{$\left.+d\delta^{2d-2}\Delta^{\frac{d}{2}-1}A_0^{-\frac{1}{2}}\numunitaries^{2d-1}\right)$}&
    $\numunitaries^{3/2}\nex^3\timeprep$ & ** &
    $2^d t^{-1}A_0^{-1}T_U$ & $\triangledown$ \\\hline
\end{tabular}
\caption{\label{tab:method_summary_extd}
Intermediate steps for the approximate calculation of the computational complexity as a function of the system size give in Tab.~\ref{tab:method_summary}. Symbols are defined in Sec.~\ref{Methods.Prelim} ($\numunitaries$, $\numterms$, $\nmeas$), Sec.~\ref{Methods.Hamsim} ($\timeunitary$), Sec.~\ref{Methods.QPEDef} ($\Ktot$, $A_0$, $t$), Sec.~\ref{Methods.ResErr} ($\Delta$) and Sec.~\ref{Methods.StatePrep} ($\timeprep$).}
\end{table*}

\begin{table}
\begin{tabular}{|c|c|c|}\hline
& Lower bound & Upper bound \\\hline
$\timeprep$ & $\nsys^2$ & $\nsys^5$ \\\hline
$\timeunitary$ & $1$ & $\nsys^5$ \\\hline
$\numterms$ & $\nsys$ & $\nsys^4$ \\\hline
$\numunitaries$ & $\nsys$ & $\nsys^4$ \\\hline
$\nex$ & $\nsys$ & $\nsys^2$ \\\hline
$t^{-1}$ & $\nsys$ & $\nsys$ \\\hline
$A_0^{-1}$ & $\nsys$ & $\nsys$ \\\hline
$\delta^{-1}$ & 1 & 1 \\\hline
$\Delta^{-1}$ & 1 & 1 \\\hline
\end{tabular}
\caption{\label{tab:approx_summary}Summary of approximations used to derive the scaling laws with $\nsys$ in Tab.~\ref{tab:method_summary}.}
\end{table}

\subsection{Quantum simulation of the electronic structure problem}

\subsubsection{Preliminaries}\label{Methods.Prelim}

To represent the electronic structure problem on a quantum computer, we need to rewrite the fermionic creation and annihilation operators $\chatdag_i$, $\chat_i$ in terms of qubit operators (e.g.~elements of the Pauli basis $\PP^N  = \{I,X,Y,Z\}^{\otimes N}$).
This is necessarily a non-local mapping, as local fermionic operators anti-commute, while qubit operators commute.
A variety of such transformations are known, including the Jordan-Wigner~\cite{jordan1928p,ortiz2001quantum}, and Bravyi-Kitaev~\cite{bravyi2002fermionic} transformations, and more recent developments~\cite{verstraete2005mapping,seeley2012bravyi,
whitfield2016local,steudtner2018fermion,
setia2018superfast,setia2018bravyi,
tranter2018comparison,obrien2018majorana}.

After a suitable qubit representation has been found, we need to design quantum circuits to implement unitary transformations.
Such circuits must be constructed from an appropriate set of primitive units, known as a (universal) gate-set.
For example, one might choose the set of all single-qubit operators, and a two-qubit entangling gate such as the controlled-NOT, C-Phase, or iSWAP gates~\cite{nielsen2002quantum}.
One can then build the unitary operators $e^{i\theta\hat{P}}$ for $\hat{P}\in\PP^N$ exactly (in the absence of experimental noise) with a number of units and time linear in the size of $\hat{P}$~\cite{whitfield2011simulation}.
(Here, size refers to the number of non-identity tensor factors of $\hat{P}$.)
Optimizing the scheduling and size of these circuits is an open area of research, but many improvements are already known~\cite{hastings2014improving}.

Transformations of a quantum state must be unitary, which is an issue if one wishes to prepare e.g. $\partial\ham/\partial\lambda|\Psi_0\rangle$ on a quantum register ($\partial\ham/\partial\lambda$ is almost always not unitary).
To circumvent this, one must decompose $\partial\ham/\partial\lambda$ as a sum of $\numunitaries$ unitary operators, perform a separate circuit for each unitary operator, and then combine the resulting measurements as appropriate.
Such a decomposition may always be performed using the Pauli group (although better choices may exist).
Each such decomposition incurs a multiplicative cost of $\numunitaries$ to the computation time, and further increases the error in any final result by at worst a factor of $\numunitaries^{1/2}$.
This makes the computational complexities reported in Tab.~\ref{tab:method_summary_extd} highly dependent on $\numunitaries$.
The scaling of $\numunitaries$ with the system size is highly dependent on the operator to be decomposed and the choice of decomposition.
When approximating this in Tab.~\ref{tab:approx_summary} we use a range between $O(\nsys)$ (which would suffice for a local potential in a lattice model) to $O(\nsys^4)$ (for a two-body interaction).

To interface with the outside world, a quantum register needs to be appropriately measured.
Similarly to unitary transformations, one builds these measurements from primitive operations, typically the measurement of a single qubit in the $Z$ basis.
This may be performed by prior unitary rotation, or by decomposing an operator $\hat{O}$ into $\numterms$ Hermitian terms $\hat{O}_i$ (which may be measured separately).
$\numterms$ differs from $\numunitaries$ defined above, as the first is for a Hermitian decomposition of a derivative operator and the second is for a unitary decomposition.
Without a priori knowledge that the system is near an eigenstate of a operator $\hat{O}$ to be measured, one must perform $\nmeas$ repeated measurements of each $\hat{O}_i$ to estimate $\langle\hat{O}\rangle$ to an accuracy $\propto \nmeas^{-1/2}\numterms^{1/2}$.
As such measurement is destructive, this process requires $\nmeas\numterms$ preparations and pre-rotations on top of the measurement time.
This makes the computational costs reported in Tab.~\ref{tab:method_summary_extd} highly dependent on $\numterms$.
The scaling of $\numterms$ with the system size $\nsys$ is highly dependent on the operator $\hat{O}$ to be measured and the choice of measurements to be made~\cite{mcclean2016theory,romero2018strategies}.
In Tab.~\ref{tab:approx_summary}, we assume a range between $O(\nsys)$ and $O(\nsys^4)$ to calculate the approximate computation cost.
This is a slight upper bound, as terms can be measured simultaneously if they commute, and error bounds may be tightened by accounting for the covariance between non-commuting terms~\cite{mcclean2016theory}.
The details on how this would improve the asymptotic scaling are still lacking in the literature however, and so we leave this as an obvious target for future work.

\begin{figure}
\includegraphics[width=\columnwidth]{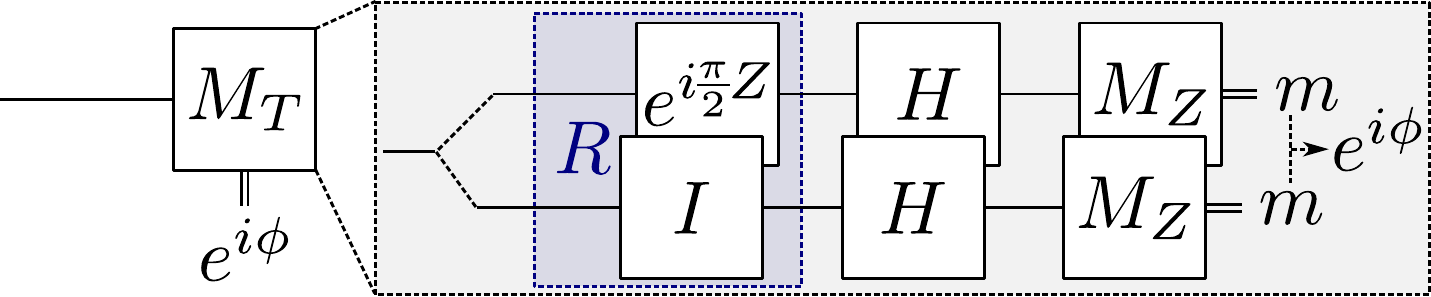}
\caption{\label{fig:tomo_circuit}Definition of the circuit element $M_T$ used throughout this work to estimate the phase $e^{i\phi}$ on a single qubit. This is done by repeatedly preparing and measuring the qubit along two different axis, by a combination of rotation and Hadamard gates and measurement $M_Z$ in the computational basis. The final measurements may then be combined via Eq.~\ref{eq:tomo_circuit_def}.}
\end{figure}

Throughout this text we require the ability to measure a phase $e^{i\phi}$ between the $|0\rangle$ and $|1\rangle$ states of a single qubit.
(This information is destroyed by a measurement in the $Z$ basis, which may only obtain the amplitude on either state.)
Let us generalize this to a mixed state on a single qubit, which has the density matrix~\cite{nielsen2002quantum}
\begin{equation}
\rho=\left(\begin{array}{cc}p_0 & p_+e^{i\phi}\\p_+e^{-i\phi} & p_1\end{array}\right),
\end{equation}
where $p_0+p_1=1$, and $0\leq p_+\leq \sqrt{p_0p_1}\leq 0.5$.
If one repeatedly performs the two circuits in Fig.~\ref{fig:tomo_circuit} (which differ by a gate $R=I$ or $R=R_Z=e^{i\frac{\pi}{4} Z}$), and estimates the probability of a final measurement $m=0,1$, one may calculate
\begin{align}
2p_+e^{i\phi}&=P(m=0|R=I)-P(m=1|R=I)\nonumber\\+&iP(m=0|R=R_Z)-iP(m=1|R=R_Z).\label{eq:tomo_circuit_def}
\end{align}
We define the circuit element $M_T$ throughout this work as the combination of the two circuits to extract a phase using this equation.
As above, the error in estimating the real and imaginary parts of $2p_+e^{i\phi}$ scales as $\nmeas^{-1/2}$.

\subsubsection{Hamiltonian Simulation}\label{Methods.Hamsim}

Optimal decompositions for complicated unitary operators are not in general known.
For the electronic structure problem, one often wants to perform time evolution by a Hamiltonian $\ham$, requiring a circuit for the unitary operator $U=e^{i\ham t}$.
For a local (fermionic or qubit) Hamiltonian, the length $\timeunitary$ of the required circuit is polynomial in the system size $\nsys$.
However, the coefficient of this polynomial is often quite large; this depends on the chosen Hamiltonian, its basis set representation, the filling factor $\fillingfactor$ (i.e.~number of particles), and whether additional ancilla qubits are used~\cite{mcardle2018quantum,cao2018quantum}.
Moreover, such circuits usually approximate the target unitary $U$ with some $\tilde{U}$ with some bounds on the error $\epsilon_{\mathrm{H}}=\|U-\tilde{U}\|_{\mathrm{S}}$.
This bound $\epsilon_{\mathrm{H}}$ is proportional to the evolution time $t$, providing a `speed limit' for such simulation~\cite{berry2007efficient}.
For the electronic structure problem, current methods achieve scaling between $O(\nsys^2)$~\cite{low2018hamiltonian} and $O(\nsys^6)$~\cite{hastings2014improving,motzoi2017linear} for the circuit length $\timeunitary$, assuming $\fillingfactor\propto\nsys$ (and fixed $t$, $\epsilon$).
(When $\fillingfactor$ is sublinear in $\nsys$, better results exist~\cite{babbush2018quantum}.)
The proven $O(\nsys^6)$ scaling is an upper bound, and most likely reduced by recent work~\cite{motta2018low,campbell2018random}.
For simpler models, such as the Hubbard model, scalings between $O(1)$ and $O(\nsys)$ are available~\cite{obrien2018majorana,kivlichan2019improved}.
As we require $t\propto\nsys^{-1}$ for the purposes of phase estimation (described in Sec.~\ref{Methods.QPEDef}), this scaling is reduced by an additional factor throughout this work (though this cannot reduce the scaling below $O(1)$).
For Tab.~\ref{tab:approx_summary}, we use a range of $\timeunitary=O(1)$ and $\timeunitary=O(\nsys^5)$ when approximating the scaling of our methods with the system size.

\subsubsection{Ground state preparation and measurement}\label{Methods.StatePrep}
A key requirement for our derivative estimation methods is the ability to prepare the ground state $|\Psi_0\rangle$ or an approximation to it on the system register.
Various methods exist for such preparation, including QPE (see Sec.~\ref{Methods.QPEDef}), adiabatic state preparation~\cite{wu2002polynomial}, VQE~\cite{peruzzo2014variational,mcclean2016theory}, and more recent developments~\cite{motta2019quantum,kyriienko2019quantum}.
Some of these preparation methods (in particular adiabatic and variational methods) are unitary, whilst others (phase estimation) are projective.
Given a unitary preparation method, one may determine whether the system remains in the ground state by inverting the unitary and measuring in the computational basis (Sec.~\ref{Methods.Prelim}).
By contrast, such determination for QPE requires another phase estimation round, either via multiple ancilla qubits or by extending the methods in Sec.~\ref{Methods.QPEDef}.
Unitary preparation methods have a slight advantage in estimating expectation values of unitary operators $\hat{U}$; the amplitude amplification algorithm~\cite{knill2007optimal} improves convergence of estimating $\langle\hat{U}\rangle$ from $\epsilon\propto T^{-1/2}$ to $\epsilon\propto T^{-1}$ (in a total computation time $T$).
However, this algorithm requires repeated application of the unitary preparation whilst maintaining coherence, which is probably not achievable in the near-term.
We list the computation time in Tabs.~\ref{tab:method_summary} and \ref{tab:method_summary_extd} both when amplitude amplification is (marked with $**$) and is not available.

Regardless of the method used, state preparation has a time cost that scales with the total system size.
For quantum phase estimation, this is the time cost $K\timeunitary$
of applying the required estimation circuits,
where $\Ktot$ is the total number of applications of $e^{i\ham t}$~\cite{higgins2009demonstrating}.
The scaling of a VQE is dependent on the variational ansatz chosen~\cite{mcclean2016theory,romero2018strategies}.
The popular UCCSD ansatz for the electronic structure problem has a $O(\nsys^5)$ computational cost if implemented naively.
However, recent work suggests aggressive truncation of the number of variational terms can reduce this as far as $O(\nsys^2)$~\cite{romero2018strategies}.
We take this as the range of scalings for our approximations in Tab.~\ref{tab:method_summary}.

\subsubsection{Systematic error from ground state approximations (state error)}\label{Methods.GSErrorSystematic}
A variational quantum eigensolver is not guaranteed to prepare the true ground state $|\Psi_0\rangle$, but instead some approximate ground state
\begin{equation}
|\tilde{\Psi}_0\rangle=\sum_j a_j|\Psi_j\rangle.
\end{equation}
In general we expect $|a_0|^2$ to be relatively large, although this may not be the case for systems with a small gap $\delta$ to nearby excited states.
One may place very loose bounds on the error this induces in the energy:
\begin{align}
2\|\ham\|_{\mathrm{S}}(1-|a_0|^2)\geq|E_0-\tilde{E}_0|&=\sum_{j>0}a_j^*a_j(E_j-E_0)\nonumber\\&\geq\delta(1-|a_0|^2)\geq 0,
\end{align}
where here $\|\ham\|_{\mathrm{S}}$ is the spectral norm of the Hamiltonian (its largest eigenvalue).
(Note that while in general $\|\ham\|_{\mathrm{S}}$ is difficult to calculate, reasonable approximations are usually obtainable.)
As $|\tilde{\Psi}_0\rangle$ is chosen to minimize the approximate energy $\tilde{E}_0$, one expects to be much closer to the smaller bound than the larger.
For an operator $\hat{D}$ (such as a derivative operator $\partial \ham/\partial\lambda$) other than the Hamiltonian, cross-terms will contribute to an additional error in the expectation value $D_0=\langle|\hat{D}|\rangle$:
\begin{align}
&|\tilde{D}_0-D_0|=\left|\sum_{i,j>0}a_i^*a_j\langle\Psi_i|\hat{D}|\Psi_j\rangle\right.\nonumber\\ &\left.+2\mathrm{Re}\sum_{j}a_j^*a_0\langle\Psi_j|\hat{D}|\Psi_0\rangle + (|a_0|^2-1)D_0\right|.
\end{align}
One can bound this above in turn using the fact that
\begin{equation}
\sum_{i,j}a_i^*a_j\leq(\sum_i|a_i|^2)^{1/2}(\sum_j|a_j|^2)^{1/2}=(1-|a_0|^2),
\end{equation}
which leads to
\begin{equation}
|\tilde{D}_0-D_0|\leq 2\|\hat{D}\|_{\mathrm{S}}\left[(1-|a_0|^2)+|a_0|\sqrt{1-|a_0|^2}\right].
\end{equation}
Combining this with the error in the energy gives the bound
\begin{equation}
|\tilde{D}_0-D_0|\leq 2\|\hat{D}\|_{\mathrm{S}}\left(\frac{|\tilde{E}_0-E_0|^{1/2}}{\delta^{1/2}}+\frac{|\tilde{E}_0-E_0|}{\delta}\right).
\end{equation}
This ties the error in our derivative to the energy in our error, but with a square root factor that unfortunately slows down the convergence when the error is small.
(This factor comes about precisely because we do not expect to be in an eigenstate of the derivative operator.)
Unlike the above energy error, we cannot expect this bound to be loose without a good reason to believe that the orthogonal component $\sum_{j>0} a_j|\Psi_j\rangle$ has a similar energy gradient to the ground state.
This will often not be the case; the low-energy excited state manifold is usually strongly coupled to the ground state by a physically-relevant excitation, causing the energies to move in opposite directions.
Finding methods to circumvent this issue are obvious targets for future research.
For example, one could optimize a VQE on a cost function other than the energy.
One could also calculate the gradient in a reduced Hilbert space (see Sec.~\ref{Methods.LR}) using eigenstates of $\ham^{(\rm QSE)}+\epsilon\hat{D}^{(\rm QSE)}$ with small $\epsilon$ to ensure the coupling is respected.\\

\subsection{Quantum Phase Estimation}\label{Methods.QPEDef}

Non-trivial measurement of a quantum computer is of similar difficulty to non-trivial unitary evolution.
Beyond learning the expectation value of a given Hamiltonian $\ham$, one often wishes to know specific eigenvalues $E_i$ (in particular for the electronic structure problem, the ground and low-excited state energies).
This may be achieved by QPE~\cite{kitaev1995quantum}.
QPE entails repeated hamiltonian simulation (as described above), conditional on an ancilla qubit prepared in the $|+\rangle=\frac{1}{\sqrt{2}}(|0\rangle+|1\rangle)$ state.
(The resource cost in making the evolution conditional is constant in the system size.)
Such evolution causes phase kickback on the ancilla qubit; if the system register was prepared in the state $\sum_j a_j|\Psi_j\rangle$, the combined (system plus ancilla) state evolves to
\begin{equation}
\sum_j a_j|\Psi_j\rangle\otimes(|0\rangle+e^{ikE_j t}|1\rangle)\label{eq:QPE_state_eq}.
\end{equation}
Repeated tomography at various $k$ allows for the eigenphases $E_j$ to be inferred, up to a phase $E_jt+2\pi\equiv E_jt$.
This inference can be performed directly with the use of multiple ancilla qubits~\cite{kitaev1995quantum}, or indirectly through classical post-processing of a single ancilla tomographed via the $M_T$ gate of Fig.~\ref{fig:tomo_circuit}~\cite{knill2007optimal,higgins2009demonstrating,
svore2013faster,kimmel2015quantum,wiebe2016efficient,
obrien2019quantum}.

The error in phase estimation comes from two sources; the error in Hamiltonian simulation and the error in the phase estimation itself.
The error in Hamiltonian simulation may be bounded by $\epsilon_{\mathrm{H}}$ (as calculated in the previous section), which in turn sets the time for a single unitary $T_U$.
Assuming a sufficiently large gap to nearby eigenvalues, the optimal scaling of the error in estimating $E_j$ is $A_j^{-1}t^{-1}\Ktot^{-1}$
(where $A_j=|a_j|^2$ is the amplitude of the $j$th eigenstate in the prepared state).
Note that the phase equivalence $E_jt+2\pi=E_jt$ sets an upper bound on $t$; in general we require $t\propto\|\ham\|_{S}$, which typically scales with $\nsys$.
(This scaling was incorporated into the estimates of $\timeunitary$ in the previous section.)
The scaling of the ground state amplitude $A_0$ with the system size is relatively unknown, although numeric bounds suggest that it scales approximately as $1-\alpha\nsys$~\cite{reiher2017elucidating}
for reasonable $\nsys$, with $\alpha$ a small constant.
Approximating this as $\nsys^{-1}$ implies that $\Ktot\propto\nsys^2$ is required to obtain a constant error in estimating the ground state energy.

The error in estimating an amplitude $A_j$ during single-ancilla QPE has not been thoroughly investigated.
A naive least-squares fit to dense estimation leads to a scaling of $\nmeas^{-1/2}k_{\max}^{-1/3}$, where $\nmeas$ is the number of experiments performed at each point $k=1,\ldots, k_{\max}$.
One requires to perform controlled time evolution for up to $k_{\max}\approx \max(t^{-1},A_j^{-1})$ coherent steps in order to guarantee separation of $\phi_j$ from other phases.
To obtain a constant error, one must then perform $\nmeas\propto k_{\max}^{-\frac{2}{3}}\propto \max(A_j^{\frac{2}{3}},t^{\frac{2}{3}})$ measurements at each $k$, implying $\Ktot=\nmeas k_{\max}\propto \max(A_j^{-\frac{1}{3}},t^{-\frac{1}{3}})$ applications of $e^{i\ham t}$ are required.
For the ground state ($A_j=A_0$), this gives scaling $\Ktot\propto\nsys^{\frac{1}{3}}$.
By contrast, multiple-ancilla QPE requires $\Nmeas$ repetitions of $e^{i\ham t}$ with $k_{\max} =\max(A_0^{-1},t^{-1})$ to estimate $A_0$ with an error of $(A_0(1-A_0)\Nmeas^{-1})^{1/2}$.
This implies that $\Nmeas\propto A_0$ measurements are sufficient, implying $\Ktot\propto\Nmeas A_0^{-1}$ may be fixed constant as a function of the system size for constant error in estimation of $A_0$.
Though this has not yet been demonstrated for single-round QPE, we expect it to be achievable and assume this scaling in this work.

\subsection{The propagator and phase estimation method}
\label{Methods.Second.QPE}

In this section, we outline the circuits required and calculate the estimation error for our newly developed PPE method for derivative estimation.

\subsubsection{Estimating expectation values with single-ancilla QPE}\label{Methods.QPEbased.expval}

\begin{figure}
\includegraphics[width=\columnwidth]{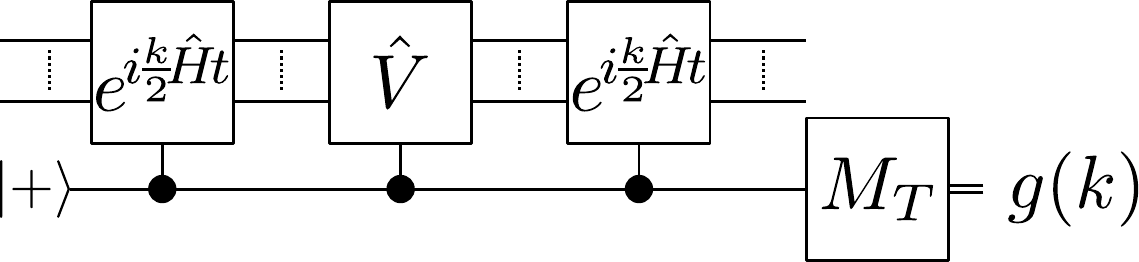}
\caption{\label{fig:Single_ancilla_QPE_circuit}A circuit to measure $\langle\Psi_j|\hat{V}|\Psi_j\rangle$ without preparing $|\Psi_j\rangle$ on the system register. The tomography box $M_T$ is defined in Fig.~\ref{fig:tomo_circuit}.}
\end{figure}

Though single-ancilla QPE only weakly measures the system register, and does not project it into an eigenstate $|\Psi_j\rangle$ of the chosen Hamiltonian, it can still be used to learn properties of the eigenstates beyond their eigenvalues $E_j$.
In particular, if one uses the same ancilla qubit to control a further unitary operation $\hat{U}$ on the system register, the combined (system plus ancilla) state evolves from Eq.~\ref{eq:QPE_state_eq} to
\begin{equation}
\sum_{j,j'} a_j(|0\rangle\otimes|\Psi_j\rangle+e^{ikE_j t}\langle \Psi_{j'}|\hat{U}|\Psi_j\rangle |1\rangle\otimes|\Psi_{j'}\rangle)\label{eq:QPE_state_eq_U}.
\end{equation}
The phase accumulated on the ancilla qubit may then be calculated to be
\begin{equation}\label{eq:gk_U}
g(k)=\sum_{j,j'} a_ja_{j'}^*\langle\Psi_{j'}|\hat{U}|\Psi_{j}\rangle e^{ik E_jt}.
\end{equation}
Note that the gauge degree of freedom is not present in Eq.~\ref{eq:gk_U}; if one re-defines $|\Psi_j\rangle\rightarrow e^{i\theta}|\Psi_j\rangle$, one must send $a_j\rightarrow e^{-i\phi_j}a_j$, and the phase cancels out.
One may obtain $g(k)$ at multiple points $k$ via tomography of the ancilla qubit (Fig.~\ref{fig:tomo_circuit}).
From here, either Prony's method or Bayesian techniques may be used to extract phases $\omega_j\approx E_jt$ and corresponding amplitudes $\alpha_j\approx \sum_{j'} a_ja_{j'}^*\langle\Psi_{j'}|\hat{U}|\Psi_{j}\rangle$~\cite{obrien2019quantum}.
The amplitudes $\alpha_j$ are often not terribly informative, but this changes if one extends this process over a family of operators $U$.
For instance, if one chooses $U=e^{ik'\ham t}\hat{V}e^{ik\ham t}$ (with $\hat{V}$ unitary), an application of Prony's method on $k$ returns amplitudes of the form
\begin{equation}
\alpha_j(k')\approx\sum_{j'}a_ja_{j'}^*e^{ik' E_{j'}t}\langle\Psi_{j'}|\hat{V}|\Psi_j\rangle,\label{eq:alphaj_def}
\end{equation}
from which a second application of Prony's method obtains phases $\omega_{j'}=E_{j'}t$ with corresponding amplitudes
\begin{equation}
\alpha_{j,j'}\approx a_ja_{j'}^*\langle\Psi_{j'}|\hat{V}|\Psi_j\rangle.\label{eq:alphajj_def}
\end{equation}
Each subsequent application of QPE requires taking data with $U^k$ fixed from $k=1,\ldots,K$ to resolve $K$ individual frequencies (and corresponding eigenvalues).
However, if one is simply interested in expectation values $\langle\Psi_j|\hat{V}|\Psi_{j}\rangle$ (i.e.~when $j=j'$), one may fix $k=k'$ and perform a single application of Prony's method, reducing the number of circuits that need to be applied from $O(K^2)$ to $O(K)$ (see Fig.~\ref{fig:Single_ancilla_QPE_circuit}).
The error in the estimator $\alpha_{j,j}$ (Eq.~\ref{eq:alphajj_def}) may be bounded above by the error in the estimator $\alpha_j$ (Eq.~\ref{eq:alphaj_def}).
However, to estimate $\langle\Psi_j|\hat{V}|\Psi_{j}\rangle$ from Eq.~\ref{eq:alphajj_def}, one needs to divide by $A_j$.
This propagates directly to the error, which then scales as $A_j^{-1/2}\Nmeas^{-1/2}$.
Thus constant error in estimating $\langle\Psi_j|\hat{V}|\Psi_{j}\rangle$ is achieved if $\Ktot\propto\nsys$.

\subsubsection{PPE circuits}\label{Methods.QPEGFCircuit}

\begin{figure*}
\includegraphics[width=\textwidth]{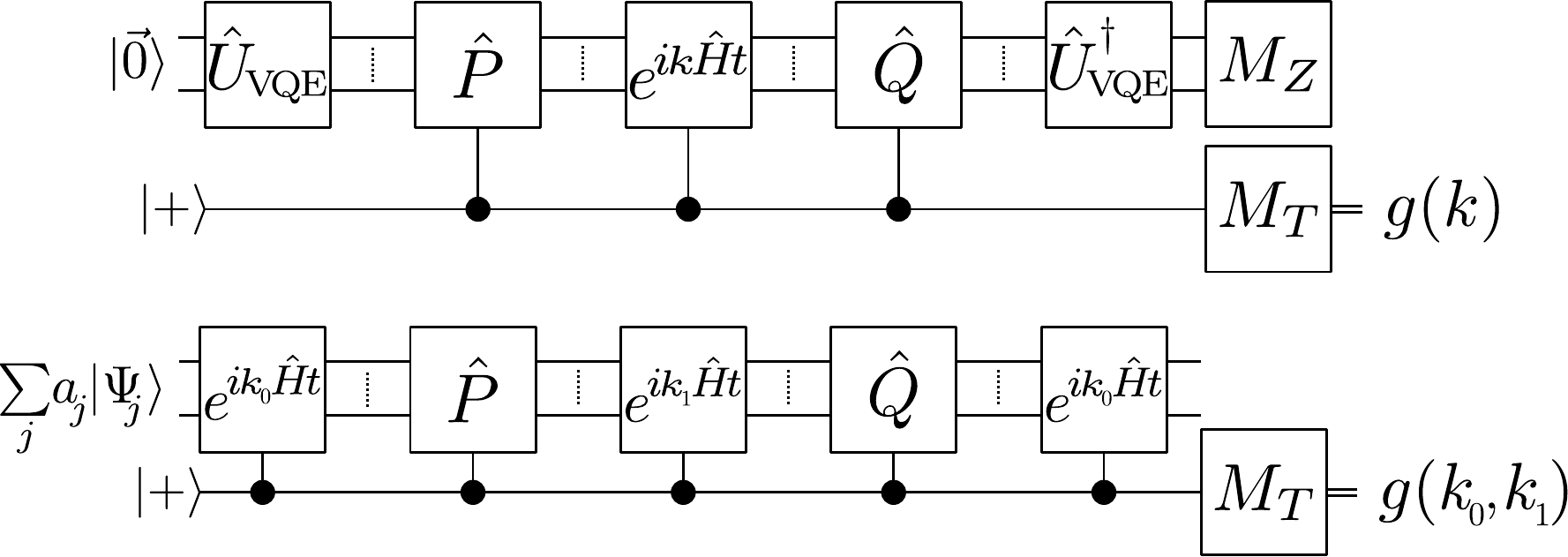}
\caption{\label{fig:GF_circuit}Circuits for calculating path amplitudes (Eq.~\ref{eq:Amplitudes}) for a second-order derivative on a quantum computer (individual units described throughout Sec.~\ref{Methods.Prelim}), using either a VQE (top) or QPE (bottom) for state preparation. Both circuits require an $N$-qubit system register and a single ancilla qubit. Repeat measurements of these circuit at different values of $k$ (top) or $k_0$ and $k_1$ (bottom) allow for the inference of the amplitude, as described in the text. $M_Z$ refers to a final measurement of all qubits in the computational basis, which is required for post-selection.}
\end{figure*}

As presented, the operator $\hat{V}$ in Fig.~\ref{fig:Single_ancilla_QPE_circuit} must be unitary.
However if one applies additional phase estimation within $\hat{V}$ itself, one can extend this calculation to non-unitary operators, such as those given in Eq.~\ref{eq:Greens_functions}.
This is similar in nature to calculating the time-domain Green's function for small $t$ on a quantum computer (which has been studied before in Refs.~\cite{wecker2015solving,dallaire2016method,bauer2016hybrid}), but performing the transformation to frequency space with Prony's method instead of a Fourier transform to obtain better spectral resolution.
It can also be considered a generalization of Ref.~\cite{roggero2018linear} beyond the linear response regime.
To calculate a $X$th order amplitude (Eq.~\ref{eq:Amplitudes}), one may set
\begin{equation}\label{eq:V_unitary}
\hat{V}=\left(\prod_{x=1}^{X-1} \hat{P}_xe^{ik_x\ham t}\right)\hat{P}_X,
\end{equation}
which is unitary if the $\hat{P}_x$ are chosen to be a unitary decomposition of $\partial\ham/\partial\lambda_x$.
In Fig.~\ref{fig:GF_circuit}, we show two circuits for the estimation of a second order derivative with $\hat{P}=\partial\ham/\partial\lambda_1$, $\hat{Q}=\partial\ham/\partial\lambda_2$ (or some piece thereof).
The circuits differ by whether QPE or a VQE is used for state preparation.
If QPE is used for state preparation, the total phase accumulated by the ancilla qubit over the circuit is
\begin{align*}
g(k_0,k_1)=\sum_{j,m,n}&a_m^*a_n\langle\Psi_m|\hat{P}|\Psi_j\rangle\langle\Psi_j|\hat{Q}|\Psi_n\rangle\\&\times e^{ik_0t(E_m+E_n)}e^{ik_1tE_j}
\end{align*}
Applying Prony's method at fixed $k_1$ will obtain a signal at phase $2tE_0$ with amplitude
\begin{equation}
\alpha_{0,0}(k_1)\approx\sum_{j}a_0^*a_0\langle\Psi_0|\hat{P}|\Psi_j\rangle\langle\Psi_j|\hat{Q}|\Psi_0\rangle e^{ik_1tE_j}
\end{equation}
A second application of Prony's method in $k_1$ allows us to obtain the required amplitudes
\begin{equation}
\alpha_{0,j,0}\approx a_0^*a_0\langle\Psi_0|\hat{P}|\Psi_j\rangle\langle\Psi_j|\hat{Q}|\Psi_0\rangle,
\end{equation}
and the eigenvalues $\omega_j\approx E_jt$, allowing classical calculation of both the amplitudes and energy coefficients required to evaluate Eq.~\ref{eq:Greens_functions}.
If a VQE is used for state preparation instead, one must post-select on the system register being returned to $|\vec{0}\rangle$.
Following this, the ancilla qubit will be in the state
\begin{equation*}
\frac{1}{\sqrt{2}}\left[|0\rangle + |1\rangle\sum_{j}e^{iktE_j}\langle\Psi_0^{(\rm VQE)}|\hat{P}|\Psi_j\rangle\langle\Psi_j|\hat{Q}|\Psi_0^{(\rm VQE)}\rangle\right],
\end{equation*}
with an accumulated phase $g(k)=\alpha_{0,0}(k)$ (where $\alpha_{0,0}$ is as defined above).
Here, $|\Psi_0^{(\rm VQE)}\rangle$ is the state prepared by the VQE unitary (which may not be the true ground state of the system).
Both methods may be extended immediately to estimate higher-order amplitudes by inserting additional excitations and rounds of QPE, resulting in amplitudes of the form $\alpha_{0,0}(k_1,\ldots,k_X)$.
To explore this in more detail, in App.~\ref{app:QPE_example} we apply this method to a simple toy system.

We note that the VQE post-selection does not constitute `throwing away data'; if the probability of post-selecting $|\Psi_0\rangle$ is $p$, we have
\begin{equation}
\sum_{k_1,\ldots,k_X}|\alpha_{0,0}(k_1,\ldots,k_X)|^2 = p,
\end{equation}
and as the variance in any term $\alpha_{i,j}(k_1,\ldots,k_X)$ scales as $|\alpha_{i,j}(k_1,\ldots,k_X)(1-\alpha_{i,j}(k_1,\ldots,k_X))|$, the absolute error in estimating a derivative scales as $p^{1/2}$ (note the lack of minus sign).
(Note here that the relative error scales as $p^{-1/2}$.)

\subsubsection{Energy discretization (resolution error)}\label{Methods.ResErr}
The maximum number of frequencies estimatable from a signal $g(0),\ldots,g(k)$ is $(k+1)/2$.
(This can be seen by parameter counting; it differs from the bound of $k$ for QPE~\cite{obrien2019quantum} as the amplitudes are not real.)
As the time required to obtain $g(k)$ scales at best linearly with $k$ (Sec.~\ref{Methods.Hamsim}), we cannot expect fine enough resolution of all $2^{\nsys}$ eigenvalues present in a $\nsys$-qubit system.
Instead, a small amplitude $\Aaa(j_1,\ldots,j_X)$ ($|\Aaa(j_1,\ldots,j_X)|\leq\Delta$) will be binned with paths $\Aaa'(l_1,\ldots,l_X)$ of similar energy ($\delta=\max_{x}|E_{j_x}-E_{l_x}|\ll \Delta$), and labeled with the same energy $E_{B_x}\approx E_{j_x}\approx E_{k_x}$~\cite{obrien2019quantum}.
Here, $\Delta$ is controlled by the length of the signal $g(k)$, i.e. $\Delta\propto\max(k)^{-1}$.
This grouping does not affect the amplitudes; as evolution by $e^{ik\ham t}$ does not mix eigenstates (regardless of energy), terms of the form $|\Psi_{j_x}\rangle\langle\Psi_{l_x}|$ do not appear.
(This additional amplitude error would occur if one attempted to calculate single amplitudes of the form $\langle\Psi_j|\hat{P}|\Psi_k\rangle$ on a quantum device, e.g.~using the method in Sec.~\ref{Methods.QPEDef} or that of Ref.~\cite{roggero2018linear}, and multiply them classically to obtain a $d>1$-th order derivative.)
The PPE method then approximates Eq.~\ref{eq:Greens_functions} as
\begin{align}
&D=\sum_{\Aaa}\sum_{B_1,\ldots,B_{X_{\Aaa}-1}}\nonumber\\&\hspace{0.5cm}\Aaa_{B_1,\ldots,B_{X_{\Aaa}-1}}f_{\Aaa}(E_0;E_{B_1},\ldots,E_{B_{X_{\Aaa}-1}}),\nonumber\\
&\Aaa_{B_1,\ldots,B_{X_{\Aaa}-1}}=\sum_{j_x\in B_x}2\;\mathrm{Re}(\Aaa(j_1,\ldots,j_{X_{\Aaa}-1}))\label{eq:approx_Greens_functions}.
\end{align}
Classical post-processing then need only sum over the (polynomially-many in $\Delta$) bins $B_x$ instead of the (exponentially-many in $\nsys$) eigenstates $|\Psi_{j_x}\rangle$, which is then tractable.

To bound the resolution error in the approximation $f_{\Aaa}(E_0;E_{j_1},\ldots,E_{j_{X_{\Aaa}-1}})\rightarrow f_{\Aaa}(E_0;E_{B_1},\ldots,E_{B_{X_{\Aaa}-1}})$, we consider the error if $E_j$ were drawn randomly from bins of width $\|\ham\|_{\mathrm{S}} \Delta$ (where $\|\ham\|_{\mathrm{S}}$ is the spectral norm).
The energy functions $f$ take the form of $X_{\Aaa}-1$ products of $\frac{1}{E_{j_x}-E_{0}}$.
If each term is independent, these may be bounded as
\begin{align}
\epsilon_{f}\leq X_{\Aaa} \delta^{(X_{\Aaa}-2)}\|\ham\|_{\mathrm{S}} \Delta,
\end{align}
where $\delta$ is the gap between the ground and excited states.
Then, as the excitations $\hat{P}$ are unitary, for each amplitude $\Aaa$ one may bound
\begin{equation}
\sum_{B_1,\ldots,B_{X_{\Aaa}-1}}|\Aaa_{B_1,\ldots,B_{X_{\Aaa}}}|^2\leq 1.
\end{equation}
Propagating variances then obtains
\begin{equation}
\epsilon_D\leq d N_{\Aaa}^{1/2}\delta^{d-2}\|\ham\|_{\mathrm{S}} \Delta,
\end{equation}
Where $N_{\Aaa}$ is the number of amplitudes in the estimation of $D$.
As we must decompose each operator into unitaries to implement in a circuit, $N_{\Aaa}\propto\numunitaries^d$.

This bound is quite loose; in general we expect excitations $\partial\ham/\partial\lambda$ to couple to low-level excited states, which lie in a constant energy window (rather than one of width $\|\ham\|_S$), and that contributions from different terms should be correlated (implying that $N_{\Aaa}$ should be treated as constant here). This implies that one may take $\Delta$ roughly constant in the system size, which we assume in this work.

\subsubsection{Sampling noise error}\label{Methods.QPEbased.sampling}

We now consider the error in calculating Eq.~\ref{eq:approx_Greens_functions} from a finite number of experiments (which is separate to the resolution error above).
Following Sec.~\ref{Methods.QPEDef} we have that, if QPE is used for state preparation
\begin{align}
&\mathrm{Var}[\Aaa_{B_1,\ldots,B_{X_{\Aaa}-1}}]\propto|\Aaa_{B_1,\ldots,B_{X_{\Aaa}-1}}|A_0^{-1}\Nmeas^{-1}\\
&\mathrm{Var}[f(E_0;E_{B_1},\ldots,E_{B_{X_{\Aaa}-1}})]\nonumber\\&\hspace{2cm}\propto X_{\Aaa} \delta^{2X_{\Aaa}-4}\Ktot^{-2}t^{-2}|\Aaa_{B_1,\ldots,B_{X_{\Aaa}-1}}|^{-2}A_0^{-2}.
\end{align}
If one were to use a VQE for state preparation, the factors of $A_0$ would be replaced by the state error of Sec.~\ref{Methods.GSErrorSystematic}.
We have not included this calculation in Tab.~\ref{tab:method_summary_extd} for the sake of simplicity.
Then, assuming each term in Eq.~\ref{eq:approx_Greens_functions} is independently estimated, we obtain
\begin{align}
&\mathrm{Var}[D]=\sum_{\Aaa}\nonumber\sum_{B_1,\ldots,B_{X_{\Aaa}}}\nonumber\\&\left\{\mathrm{Var}[f_{\Aaa}(E_0;E_{B_1},\ldots,E_{B_{X_{\Aaa}}})]\left|\Aaa_{B_1,\ldots,B_{X_{\Aaa}}}\right|^2\right.\nonumber\\
&\left.+\mathrm{Var}[\Aaa_{B_1,\ldots,B_{X_{\Aaa}}}]\left|f_{\Aaa}(E_0;E_{B_1},E_{B_2},\ldots,E_{B_{X_{\Aaa}}})\right|^2\right\}.
\end{align}
Substituting the individual scaling laws one obtains
\begin{align}
&\mathrm{Var}[D]\propto \sum_{\Aaa}\sum_{B_1,\ldots,B_{X_{\Aaa}}}\nonumber\left\{X\delta^{2d-4}K^{-2}t^{-2}A_0^{-2}\right.\nonumber\\&\hspace{2.5cm}\left.+\delta^{2d-2}|\Aaa_{B_1,\ldots,B_{X_{\Aaa}}}|A_0^{-1}\Nmeas^{-1}\right\}\\
&\leq N_{\Aaa}\left\{\delta^{2d-2}\Delta^{\frac{d}{2}}\Nmeas^{-1}A_0^{-1}+d\delta^{2d-4}\Delta^dK^{-2}t^{-2}A_0^{-2}\right\},
\end{align}
where again $N_{\Aaa}\propto N_U^d$.
This result is reported in Tab.~\ref{tab:method_summary_extd}.

\subsection{Eigenstate truncation approximation details}\label{Methods.LR}
In this section, we outline the classical post-processing required to evaluate Eq.~\ref{eq:Greens_functions} in the ETA, using QSE to generate approximate eigenstates. We then calculate the complexity cost of such estimation, and discuss the systematic error in an arbitrary response approximation from Hilbert space truncation.

The chosen set of approximate excited states $|\tilde{\Psi}_j\rangle$ defines a subspace $\Hh^{(\mathrm{QSE})}$ of the larger FCI Hilbert space $\Hh^{(\mathrm{FCI})}$.
To calculate expectation values within this subspace, we project the operators $\hat{O}$ of interest (such as derivatives like $\partial\ham/\partial\lambda$) onto $\Hh^{(\mathrm{QSE})}$, giving a set of reduced operators $\hat{O}^{(\mathrm{QSE})}$ ($O^{(\mathrm{QSE})}_{i,j}=\langle\tilde{\Psi}_i|\hat{O}|\tilde{\Psi}_j\rangle$).
These are $\nex\times\nex$-dimensional classical matrices, which may be stored and operated on in polynomial time.
Methods to obtain the matrix elements $O^{(\mathrm{QSE})}_{i,j}$ are dependent on the form of the $|\tilde{\Psi}_j\rangle$ chosen.
Within the QSE, one can obtain these by directly measuring~\cite{mcclean2017hybrid}
\begin{equation}
\langle\chi_i|\hat{O}|\chi_j\rangle=\langle\Psi_0|\hat{E}_i^{\dag}\hat{O}\hat{E}_j|\Psi_0\rangle,\label{eq:QSE_term}
\end{equation}
using the techniques outlined in Sec.~\ref{Methods.Prelim}, and rotating the basis from $\{|\chi_j\rangle\}$ to $\{|\tilde{\Psi}_j\rangle\}$ (using Eq.~\ref{eq:gen_eval}).

The computational complexity for a derivative calculation within the QSE is roughly independent of the choice of $|\tilde{\Psi}_j\rangle$.
The error $\epsilon$ may be bounded above by the error in each term of the $\nex\times\nex$ projected matrices, which scales as either $\numterms^{1/2}\nmeas^{-1/2}$ (when directly estimating), $A_j^{-1/2}\Nmeas^{-1/2}$ (when estimating via QPE), or $\numunitaries^{1/2}K^{-1}\nmeas^{-1/2}$ (using the amplitude estimation algorithm).
We assume that the $\nex^2$ terms are independently estimated, in which case $\epsilon$ scales with $\nex$.
In general this will not be the case, and $\epsilon$ could scale as badly as $\nex^2$, but we do not expect this to be typical.
Indeed, one can potentially use the covariance between different matrix elements to improve the error bound~\cite{mcclean2017hybrid}.
As we do not know the precise improvement this will provide, we leave any potential reduction in the computational complexity stated in Tab.~\ref{tab:method_summary_extd} to future work.
The calculation requires $\nmeas$ repetitions of $\numterms$ circuits for each pair of $\nex$ excitations, leading to a total number of $\nmeas\numterms\nex^2$ preparations (each of which has a time cost $\timeprep$), as stated in Tab~\ref{tab:method_summary_extd}.
(With the amplitude amplification algorithm, the dominant time cost comes from running $O(\numterms\nex^2)$ circuits of length $K\timeprep$.)

Regardless of the method of generating eigenstates, the ETA incurs a systematic truncation error from approximating an exponentially large number of true eigenstates $|\Psi_j\rangle$ by a polynomial number of approximate eigenstates $|\tilde{\Psi}_j\rangle=\sum_l\tilde{A}_{j,l}|\Psi_l\rangle$.
This truncation error can be split into three pieces.
Firstly, an excitation $\hat{P}|\Psi_0\rangle$ may not lie within the response subspace $\Hh^{(\mathrm{QSE})}$, in which case the pieces lying outside the space will be truncated away.
Secondly, the term $\hat{P}|\tilde{\Psi}_j\rangle\langle\tilde{\Psi}_j|\hat{Q}$ may contain terms of the form $\hat{P}|\Psi_j\rangle\langle\Psi_l|\hat{Q}$, which do not appear in the original resolution of the identity.
Thirdly, the approximate energies $\tilde{E}_j$ may not be close to the true energies $E_j$ (especially when $|\tilde{\Psi}_j\rangle$ is a sum of true eigenstates $|\Psi_l\rangle$ with large energy separation $E_j-E_l$).
If one chooses excitation operators $\hat{E}_j$ in the QSE so that $\hat{P}=\sum_jp_j\hat{E}_j$, one completely avoids the first error source.
By contrast, if one chooses a truncated set of true eigenstates $|\tilde{\Psi}_j\rangle=|\Psi_j\rangle$, one avoids the second and third error sources exactly.
In App.~\ref{App.ResponseBounds} we expand on this point, and place some loose bounds on these error sources.

\subsection{Experimental methods}\label{Methods.expt}

The experimental implementation of the geometry optimization algorithm was performed using two of three transmon qubits in a circuit QED quantum processor.
This is the same device used in Ref.~\cite{sagastizabal2019error} (raw data is the same as in Fig.1(e) of this paper, but with heavy subsequent processing).
The two qubits have individual microwave lines for single-qubit gating and flux-bias lines for frequency control, and dedicated readout resonators with a common feedline.
Individual qubits are addressed in readout via frequency multiplexing.
The two qubits are connected via a common bus resonator that is used to achieve an exchange gate,
\begin{equation}
\left(\begin{array}{cccc}1 & 0 & 0 & 0 \\ 0 & \cos(\theta) & i\sin(\theta) & 0 \\ 0 & i\sin(\theta) & \cos(\theta) & 0 \\ 0 & 0 & 0 & 1 \end{array}\right),
\end{equation}
via a flux pulse on the high-frequency qubit, with an uncontrolled additional single-qubit phase that was cancelled out in post-processing.
The exchange angle $\theta$ may be fixed to a $\pi/6000$ resolution by using the pulse duration (with a $1~\mathrm{ns}$ duration) as a rough knob and fine-tuning with the pulse amplitude.
Repeat preparation and measurement of the state generated by exciting to $|01\rangle$ and exchanging through one of $41$ different choices of $\theta$ resulted in the estimation of $41$ two-qubit density matrices $\rho_i$ via linear inversion tomography of $10^4$ single-shot measurements per pre-rotation~\cite{saira2014entanglement}.
All circuits were executed in eQASM~\cite{Fu2019eQASM} code compiled with the QuTech OpenQL compiler, with measurements performed using the qCoDeS~\cite{QCoDeS16} and PycQED~\cite{PycQED16} packages.

To use the experimental data to perform geometry optimization for H$_2$, the ground state was estimated via a VQE~\cite{peruzzo2014variational,mcclean2016theory}.
The Hamiltonian at a given H-H bond distance $R_{\rm H-H}$ was calculated in the STO-3G basis using the Dirac package~\cite{DIRAC18}, and converted to a qubit representation using the Bravyi-Kitaev transformation, and reduced to two qubits via exact block-diagonalization~\cite{omalley2016scalable} using the Openfermion package~\cite{mcclean2017openfermion} and the Openfermion-Dirac interface~\cite{openfermion_dirac}.
With the Hamiltonian $\ham(R_{\rm H-H})$ fixed, the ground state was chosen variationally: $\rho(R_{\rm H-H})=\min_{\rho_i}\mathrm{Trace}[\ham(R_{\rm H-H})\rho_i]$.
The gradient and Hessian were then calculated from $\rho(R_{\rm H-H})$ using the Hellmann--Feynman theorem (Sec.~\ref{Main.Scalings}) and ETA (Sec.~\ref{Methods.LR}) respectively.
For the ETA, we generated eigenstates using the QSE, with the Pauli operator $XY$ as a single excitation.
This acts within the number conserving subspace of the two-qubit Hilbert space, and, being imaginary, will not have the real-valued H$_2$ ground state as an eigenstate.
(This in turn guarantees the generated excited state is linearly independent of the ground state.)
For future work, one would want to optimize the choice of $\theta$ at each distance $R_{\rm H-H}$, however this was not performed due to time constraints.
We have also not implemented the error mitigation strategies studied in Ref.~\cite{sagastizabal2019error} for the sake of simplicity.

\subsection{Simulation methods}\label{Methods.quantumsim}
Classical simulations of the quantum device were performed in the full-density-matrix simulator \emph(quantumsim)~\cite{obrien2017density}.
A realistic error model of the device was built using experimentally calibrated parameters to account for qubit decay ($T_1$), pure dephasing ($T_2^{*}$), residual excitations of both qubits, and additional dephasing of qubits fluxed away from the sweet spot (which reduces $T_2^{*}$ to $T_{2}^{*,red}$ for the duration of the flux pulse).
This error model further accounted for differences in the observed noise model on the individual qubits, as well as fluctuations in coherence times and residual excitation numbers.
Further details of the error model may be found in Ref.~\cite{sagastizabal2019error} (with device parameters in Tab.S1 of this reference).

With the error model given, $100$ simulated experiments were performed at each of the $41$ experimental angles given.
Each experiment used unique coherence time and residual excitation values (drawn from a distribution of the observed experimental fluctuations), and had appropriate levels of sampling noise added.
These density matrices were then resampled $100$ times for each simulation.

\section{Author contributions}
New quantum algorithms were designed by TEO. Errors were estimated by TEO, BS, and AD. Experiment was performed by RS and LDC. Density matrix simulations were performed by XBM. Computational chemistry calculations performed by BS and LV. Gradient descent and polarizability calculations from experimental and simulated data were performed by TEO and BS. Relevance to quantum chemistry and suggested applications provided by FB and LV. Paper written by TEO, BS, FB, and LV (with input from all other authors).

\section{Competing interests}
The authors declare no conflicts of interest.

\begin{acknowledgments}
We would like to thank B.M. Terhal, I. Kassal, V.P. Ostroukh and C.H. Price for useful discussions, M. Singh, M.A. Rol, C.C. Bultink, X. Fu, N. Muthusubramanian, A. Bruno, M. Beekman, N. Haider, F. Luthi, B. Tarasinski and C. Dickel for experimental assistance, and C.W.J. Beenakker and D. Hohl for support during this project.
This research was funded by the Netherlands Organization for Scientific Research (NWO/OCW), an ERC Synergy Grant, Shell Global Solutions BV, and IARPA (U.S. Army Research Office grant W911NF-16-1-0071).
\end{acknowledgments}

\bibliographystyle{apsrev4-1}

\newcommand{\Aa}[0]{Aa}

\appendix

\section{Analytical derivative of the one- and two-electron integrals}\label{Methods.classical_calc.1}

Molecular orbitals are usually obtained as a linear combination of atomic orbitals (LCAO) via HF or another self-consistent field calculation:
\begin{eqnarray}\label{eq:LCAO}
\phi_p(\bfr) = \sum_{\mu}^{AO} c_{\mu p} \chi_\mu(\bfr),
\end{eqnarray}
where $\lbrace c_{\mu p} \rbrace$ are the MO coefficients and $\lbrace \chi_\mu(\bfr) \rbrace$ the atomic orbitals (AO).
The atomic orbitals are usually chosen to be Gaussian functions centered at the nuclear positions $\bfR_i$ and will move with the nuclei if the geometry is modified.
An alternative is to express $\{\phi_p(\bfr)\}$ in terms of plane-wave type or Gausslet-type basis functions~\cite{babbush2017low,
white2017hybrid}, which reduce the position dependence or reduce the number of non-zero $g_{pqrs}$ terms, respectively.

With  truncated atom-centered basis sets, one needs to consider the so-called `wavefunction forces' (see Ref.~\cite{pulay1977direct} and references therein) resulting from the basis-set dependence on the nuclear geometry. 
However, with an appropriately basis-dependent second-quantized Hamiltonian, all changes in the basis can be incorporated into those of the Hamiltonian~\cite{surjan1988second,surjan1989second,helgaker2014molecular}).

Similarly to classical computing, a quantum computer will be limited by the systematic error in the truncated basis set chosen for the problem.
Reducing this by directly increasing the number of basis functions is costly (as this defines the system size $\nsys$). Indeed,
the errors in correlation energies decay slowly as $\epsilon = O(\nsys^{-1})$ due to the presence of the Coulomb cusp in the wavefunction~\cite{helgaker2000perspective}.
To bypass this slow convergence,
the interelectron distance $r_{12}$ can be incorporated
explicitly in the wavefunction (such as R12/F12 wavefunctions), thus leading to the so-called explicitly correlated methods~\cite{klopper2006r12,kong2011explicitly}.
In particular for quantum computers, the basis set error must be traded against the number of terms $\nham$ in the Hamiltonian, and the length of the circuits required to prepare ground states or perform phase estimation; optimizing this trade-off is a topic of much debate in the field~\cite{kivlichan2019improved,babbush2017low,
poulin2017fast,berry2018improved}.

Let us consider the general case in which both the MO coefficients and the Hamiltonian matrix elements depend on the perturbation, and consider first the analytical calculation of these derivatives with respect to a general perturbation $\lambda$.
(When $\lambda=\bfR_i$ this yields the Hessian for geometry optimization, and when $\lambda=F_i$ this describes an applied electric field for a polarizability calculation.)
According to Eq.~(\ref{eq:LCAO}), the one- and two-electron integrals in the MO basis read:
\begin{eqnarray}\label{h_pq_MO_AO}
h_{pq} = \sum_{\mu \nu}^{AO} c_{\mu p}^{*} c_{\nu q}  h_{\mu \nu},
\end{eqnarray}
and
\begin{eqnarray}\label{pqrs_munurhotau}
g_{pqrs} = (pq|rs) = \sum_{\mu \nu \rho \tau}^{AO} c_{\mu p}^{*} c_{\nu q} c_{\rho r}^{*} c_{\tau s} (\mu \nu | \rho \tau ),
\end{eqnarray}
respectively.
Differentiating the above expressions by $\lambda$ leads to
\begin{eqnarray}\label{h_pq_deriv_1}
\dfrac{\partial h_{pq}}{\partial \lambda} = \sum_{\mu \nu}^{AO} \left( 
\dfrac{\partial c_{\mu p}^{*}}{\partial \lambda} c_{\nu q} h_{\mu\nu}
+ c_{\mu p}^{*} \dfrac{\partial c_{\nu q}}{\partial \lambda} h_{\mu \nu} + c_{\mu p}^{*} 
c_{\nu q} \dfrac{\partial h_{\mu \nu}}{\partial \lambda} \right),\nonumber \\
\end{eqnarray}
and
\begin{eqnarray}\label{g_pqrs_deriv_1}
&&\dfrac{\partial g_{pqrs}}{\partial \lambda}
 =  \sum_{\mu \nu \rho \tau}^{AO} \Bigg(
\dfrac{\partial c_{\mu p}^{*}}{\partial \lambda}c_{\nu q} c_{\rho r}^{*} c_{\tau s} (\mu \nu | \rho \tau ) \nonumber \\
&&+ c_{\mu p}^{*} \dfrac{\partial c_{\nu q}}{\partial \lambda} c_{\rho r}^{*} c_{\tau s} (\mu \nu | \rho \tau )
+ c_{\mu p}^{*} c_{\nu q} \dfrac{\partial c_{\rho r}^{*}}{\partial \lambda} c_{\tau s} (\mu \nu | \rho \tau )\nonumber \\
&&+ c_{\mu p}^{*} c_{\nu q} c_{\rho r}^{*} \dfrac{\partial c_{\tau s}}{\partial \lambda} (\mu \nu | \rho \tau )
+ c_{\mu p}^{*} c_{\nu q} c_{\rho r}^{*} c_{\tau s} \dfrac{\partial 
(\mu \nu | \rho \tau )}{\partial \lambda}\Bigg),\nonumber \\
\end{eqnarray}
where
\begin{eqnarray}\label{U_coeff}
\dfrac{\partial c_{\mu p}}{\partial \lambda} = \sum_{m}^{MO} c_{\mu m} U_{mp}^{(1)}.
\end{eqnarray}
The matrix $\mathbf{U}^{(1)}(\lambda)$ parametrizes the first-order changes in the MO coefficients and can be obtained by solving the coupled perturbed Hartree--Fock (CPHF) equations~\cite{gerratt1968force,caves1969perturbed,osamura1986second}.
When the number of perturbations that need to be evaluated is large, the use of an explicitly $\lambda$-dependent $\mathbf{U}$ matrix in the evaluation of the energy derivative can be avoided with the Z-vector technique of Handy and Schaeffer~\cite{handy1984evaluation}.
The last terms in Eqs.~(\ref{h_pq_deriv_1}) and (\ref{g_pqrs_deriv_1}) depend on the derivatives of the one- and two-electron integrals in the AO basis.
Those are simply given by
\begin{eqnarray}
\dfrac{\partial h_{\mu \nu}}{\partial \lambda}= \bra{ \dfrac{\partial \chi_\mu}{\partial \lambda} }
\ham \ket{\chi_\nu}
 + \bra{\chi_\mu } 
\ham \ket{ \dfrac{\partial \chi_\nu}{\partial \lambda}} + \bra{\chi_\mu } 
\dfrac{\partial \ham}{\partial \lambda} \ket{\chi_\nu}, \nonumber \\
\end{eqnarray}
and
\begin{eqnarray}
\dfrac{\partial (\mu \nu | \rho \tau)}{\partial \lambda}
& =&  ( \dfrac{\partial \chi_\mu}{\partial \lambda}
\chi_\nu | \chi_\rho \chi_\tau ) + ( \chi_\mu \dfrac{\partial \chi_\nu}{\partial \lambda} |
 \chi_\rho \chi_\tau )  \nonumber \\
 && + (\chi_\mu \chi_\nu | \dfrac{\partial \chi_\rho}{\partial \lambda}
 \chi_\tau ) + ( \chi_\mu \chi_\nu | \chi_\rho
  \dfrac{\partial \chi_\tau}{\partial \lambda}),
\end{eqnarray}
respectively.
Once all derivatives in Eqs.~(\ref{h_pq_deriv_1}) and (\ref{g_pqrs_deriv_1}) are determined, the Hamiltonian
\begin{eqnarray}\label{eq:dH_dlambda_classical}
\dfrac{\partial \ham}{\partial \lambda} = \sum_{pq} \dfrac{\partial 
h_{pq}}{\partial \lambda} \hat{E}_{pq} + \dfrac{1}{2} \sum_{pqrs} 
\dfrac{\partial g_{pqrs}}{\partial \lambda} \left( \hat{E}_{pq} \hat{E}_{rs} 
- \delta_{qr} \hat{E}_{ps} \right), \nonumber \\
\end{eqnarray}
can be mapped onto a qubit Hamiltonian via whichever mapping was chosen for said Hamiltonian (see Methods section of main text), as these encodings are Hamiltonian- (and thus perturbation-) independent.
In some cases (for instance, the application of an electric field) the derivatives of the basis functions $\lbrace \chi_\mu (\bfr) \rbrace$ are equal to zero because the basis functions do not depend on the perturbation.

\section{Numerical optimization and approximate Hessian calculations}\label{Methods.HFApprox}

Numerous numerical methods for geometry optimization exist, some gradient-free, some requiring only gradient calculations, and some making use of both gradients and Hessian data~\cite{nocedal2006numerical,schlegel2011geometry}.
As sampling noise from a quantum computer is typically far larger than the fixed point error on a classical computer, optimization techniques are required to be stable in the presence of this noise.
In particular, common implementations of algorithms that numerically estimate gradients tend to construct approximate derivatives by difference approximations, which (as we investigated above) dramatically enhance sampling noise unless care is taken.
The Nelder--Mead gradient-free algorithm~\cite{nelder1965simplex} is a common choice for optimization in quantum algorithms for this reason; as it does not rely on such an approximation, and implementations in scipy~\cite{scipy} prove relatively stable.
Gradient- and Hessian-requiring algorithms do not tend to suffer from such instability as gradient-free methods.

In this work, our geometry optimization was reduced to a one-dimensional problem, removing some of the complexity of the task.
With more atoms, one need to choose both the direction and the distance to step towards the minima of the energy landscape.
Both the CG and Newton's methods are adjustments to the steepest descent algorithm (which aims to go solely in the direction of the derivative) to account for local curvature.
In the absence of any higher order derivatives to assist adjustment, the non-linear CG algorithm weights each direction against traveling in previously-explored directions, and then performs a line-search in this direction (absent additional information that allows an estimation of how far to initially travel).
Newton's method, by comparison, benefits from access to the Hessian, allowing us to choose
\begin{equation}
\delta \bfR=\left[\frac{\partial^2E_0}{\partial\bfR^2}\right]^{-1}\frac{\partial E_0}{\partial \bfR},
\end{equation}
for the direction.
One must compensate here for the fact that we wish to minimize, and not maximise, the energy.
For our one-dimensional problem this is achieved by taking the absolute value of $\partial^2E_0/\partial R_{\mathrm{H-H}}^2$; for a higher-dimensional problem this is slightly more involved~\cite{nocedal2006numerical}.
Regardless, such modified Newton's methods tend to provide a far more optimal method for estimating higher dimensional functions than Hessian-free methods~\cite{nocedal2006numerical,schlegel2011geometry}.
We are further able to bound the minimum bond length in our geometry optimization (in particular to $R_{\mathrm{H-H}}>0.3$ \r{A}), which can be of importance for stability as classical methods tend to fail when atoms are unrealistically close together.

For large systems when low accuracy is needed (e.g.~at the start of a geometry optimization calculation), 
one may consider calculating the Hessian via the HF Hamiltonian for the same geometry as a low-cost alternative to explicit calculation on the quantum computer.
This is a standard technique for geometry optimization in computational chemistry~\cite{fischer1992general}.
As the Hessian is not used to determine convergence (which depends instead on the size of the gradient), the approximation only affects the convergence rate and stability, rather than the final result.
This is even more so for quasi-Newton methods, as the Hessian is updated during the geometry optimization by the estimated gradients, which are more accurate.
Calculating the HF Hessian is a standard procedure in most computational programs; for further mathematical information, we refer the reader to Ref.~\cite{yamaguchi1994analytic}.
This information is additionally obtained `for free' if the $\mathbf{U}^{(1)}$ matrix, required for first-order derivative calculations of the electronic integrals [Eqs.~(\ref{h_pq_deriv_1}) and (\ref{g_pqrs_deriv_1})], has been obtained (and vice-versa).

\section{Application of PPE method to toy system}\label{app:QPE_example}

\begin{figure*}
\includegraphics[width=\textwidth]{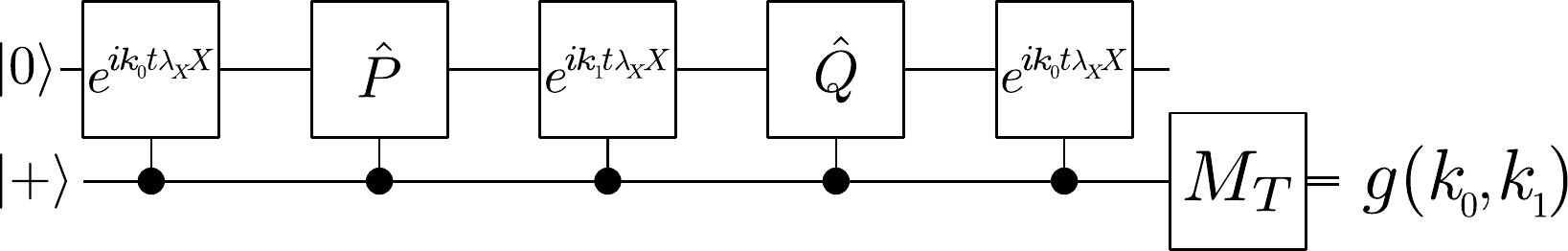}
\caption{\label{fig:toy_circuit}Circuit to perform PPE for second-order gradients on the toy system of Eq.~\ref{eq:toy_ham}, with $\lambda_Z=0$.}
\end{figure*}

In this section, we detail an example of PPE for a toy Hamiltonian,
\begin{equation}
\ham=\lambda_X X+\lambda_Z Z.\label{eq:toy_ham}.
\end{equation}
One may immediately calculate eigenvalues and derivatives of this operator to second order
\begin{align}
&E_{\pm}=\pm\sqrt{\lambda_X^2+\lambda_Z^2},\\
&\frac{\partial^2 E_{\pm}}{\partial\lambda_X^2}=\frac{\lambda_Z^2}{E_{\pm}^3},\hspace{0.5cm}\frac{\partial^2 E_{\pm}}{\partial\lambda_Z^2}=\frac{\lambda_X^2}{E_{\pm}^3},\\
&\frac{\partial^2 E_{\pm}}{\partial\lambda_X\partial\lambda_Z}=-\frac{\lambda_X\lambda_Z}{E_{\pm}^3},
\end{align}
where the energy of the ground (excited) state $|E_-\rangle$ ($|E_+\rangle$) is $E_-$ ($E_+$).
We wish to recover these second-order excitations via repeated QPE (using also QPE for state preparation).
For simplicity, let us consider calculating the derivatives when $\lambda_Z=0$, which makes $|0\rangle$ an equal superposition of the two eigenstates $|E_{\pm}\rangle=\frac{1}{\sqrt{2}}(|0\rangle\pm|1\rangle)$.
The required circuit for second-order gradient calculation, using $|0\rangle$ as a starting state, is given in Fig.~\ref{fig:toy_circuit}.
Here, $\hat{P}$ and $\hat{Q}$ should be set to $X$ or $Z$ depending on which derivative is to be estimated.
Let us first consider estimating $\frac{\partial^2 E_0}{\partial\lambda_Z^2}$, which requires that we set $\hat{P}=\hat{Q}=Z$.
One can calculate the system state prior to measurement of the phase of the ancilla qubit to be
\begin{align}
\frac{1}{\sqrt{2}}\left\{\vphantom{\frac{1}{2}}|00\rangle + |1\rangle \Big[\cos(2k_0\lambda_X t-k_1\lambda_Xt)|0\rangle\right.\nonumber\\+\left.\vphantom{\frac{1}{2}}i\sin(2k_0\lambda_Xt-k_1\lambda_Xt)|1\rangle \Big]\right\},
\end{align}
and so the reduced density matrix on the ancilla qubit may be found to be
\begin{equation*}
\rho_{\mathrm{a}}=\frac{1}{2}\left(\begin{array}{cc}
1 & \cos(2k_0\lambda_Xt-k_1\lambda_Xt) \\ \cos(2k_0\lambda_Xt-k_1\lambda_Xt) & 1
\end{array}\right).
\end{equation*}
The phase measurement $M_T$ then obtains a function
\begin{align*}
g(k_0,k_1)&=\cos(2k_0\lambda_Xt-k_1\lambda_Xt)\\
&=\frac{1}{2}(e^{i(2k_0\lambda_Xt-k_1\lambda_Xt)}+e^{i(k_1\lambda_Xt-2k_0\lambda_Xt)}).
\end{align*}
A Fourier transform~\footnote{In practice, one will obtain faster convergence here with Prony's method than a Fourier transform, but the latter is probably more familiar to the reader.} in $k_0$ at fixed $k_1$ will return peaks at $2t\omega_{\pm}=\pm2t\lambda_X$, with complex amplitude (following the notation of Eq.~32 of the main text)
\begin{equation}
\alpha_{\pm,\pm}(k_1)= \frac{1}{2}e^{\mp ik_1\lambda_X t}.
\end{equation}
We further Fourier transform $\alpha_{-,-}(k_1)$ (in terms of $k_1$), as we are interested in the derivative of the ground state.
This obtains a peak at $t\omega_+= t\lambda_X$, with amplitude
\begin{equation}
\alpha_{-,+,-}= a_{-}^*a_{-}\langle E_-|Z|E_+\rangle\langle E_+|Z|E_-\rangle=\frac{1}{2}.
\end{equation}
To finish the computation of the gradient, we note that $|\alpha_{-,-}(k_1=0)|=a_{-}^*a_{-}=\frac{1}{2}$, allowing us to compute $\langle E_-|Z|E_+\rangle\langle E_+|Z|E_-\rangle=1$.
We then substitute into Eq.~5 of the main text.
\begin{equation}
\left.\frac{\partial^2 E_-}{\partial\lambda_Z^2}\right|_{\lambda_Z=0}=2\frac{\langle E_-|Z|E_+\rangle\langle E_+|Z|E_-\rangle}{\omega_--\omega_+}=\frac{-1}{\lambda_X},
\end{equation}
as required.
We observe that the above procedure just as easily obtains the gradient of the excited state $|E_+\rangle$, as our starting state was a superposition of both ground and excited states.

Let us now repeat the calculation for the other second order derivatives.
Running the circuit of Fig.~\ref{fig:toy_circuit} with $\hat{P}=X$, $\hat{Q}=Z$, the system state evolves to
\begin{equation*}
\frac{1}{\sqrt{2}}\left\{\vphantom{\frac{1}{2}}|00\rangle + |1\rangle\Big[-\cos(k_1\lambda_Xt)|1\rangle+i\sin(k_1\lambda_Xt)|0\rangle \Big]\right\},
\end{equation*}
and the total phase accumulated on the ancilla qubit may be calculated to be
\begin{equation*}
g(k_0,k_1)=i\sin(k_1t\lambda_X)=\frac{1}{2}(e^{ik_1t\lambda_X}-e^{-ik_1t\lambda_X}).
\end{equation*}
A Fourier transform of this function in $k_0$ obtains spurious peaks at 
\begin{equation}
\omega_s=\frac{1}{2}(\omega_++\omega_-)=0,
\end{equation}
but none at $\omega_-$ or $\omega_+$, implying all amplitudes are zero, and the total derivative $\left.\frac{\partial^2 E_-}{\partial\lambda_X\partial\lambda_Z}\right|_{\lambda_Z=0}$ is $0$ as well.

Finally, running the circuit of Fig.~\ref{fig:toy_circuit} with $\hat{P}=\hat{Q}=X$ gives a pre-measurement state of
\begin{align*}
\frac{1}{\sqrt{2}}\left\{\vphantom{\frac{1}{2}}|00\rangle + |1\rangle[-\cos(2k_0\lambda_Xt+k_1\lambda_Xt)|0\rangle\right.\nonumber\\-\left.\vphantom{\frac{1}{2}}i\sin(2k_0\lambda_Xt+k_1\lambda_Xt)|1\rangle]\right\},
\end{align*}
and an accumulated phase of
\begin{align*}
g(k_0,k_1)&=-\cos(2k_0\lambda_Xt+k_1\lambda_Xt)\\&=\frac{-1}{2}(e^{i(2k_0\lambda_Xt+k_1\lambda_Xt)}+e^{i(-2k_0\lambda_Xt-k_1\lambda_Xt)}).
\end{align*}
This time, a Fourier transform at fixed $k_1$ obtains a peak at $\omega_-=-t\lambda_X$ as required, but with an amplitude
\begin{equation}
\alpha_{-,-}(k_1)=\frac{-1}{2}e^{-ik_1\lambda_Xt}.
\end{equation}
The second Fourier transform in $k_1$ will then return a peak at $\omega_-$ again (and no peak at $\omega_+$).
One must then note that the sum over eigenstates in Eq.~5 of the main text is over excited states only, implying that this peak should be excluded, and the final derivative is
\begin{equation}
\left.\frac{\partial^2 E_-}{\partial\lambda_Z^2}\right|_{\lambda_Z=0}=0,
\end{equation}
as required.

The considerations in calculating the last two above derivatives demonstrate facets of the PPE resolution error.
In practice, one must take sufficient data to resolve any spurious peaks from the true ground state (so as to not confuse these contributions).
In a larger system, spurious peaks may appear halfway between any pair of eigenstates, corresponding to scattering processes that do not return the system to the initial state.
Interestingly, these peaks will be absent if a VQE is used for state preparation, which suggests the possibility that they might not be fundamental to the estimation technique.
One must also take sufficient data to determine whether peaks from the second Fourier transform correspond to excitations from the ground state to the ground state, so that they may be excluded.
Both such methods then require a resolution on the order of the gap between ground and excited states.

\section{Bounding the systematic error in the ETA}\label{App.ResponseBounds}
In this section we attempt to expand on the sources of systematic error in the ETA.
This approximation revolves around the approximation of the exponentially-many true eigenstates $|\Psi_j\rangle$ with a set of polynomially-many approximate eigenstates $|\tilde{\Psi}_j\rangle$.
Let us write this transformation in the eigenbasis of the original Hamiltonian $\ham$, as the product of a projection $\Pi^{(\mathrm{ETA})}$ into the Hilbert space $\Hh^{(\mathrm{ETA})}$, followed by a rotation $V^{(\mathrm{ETA})}$ into the eigenbasis of the projected Hamiltonian
\begin{equation}
\Pi^{(\mathrm{ETA})}\ham\Pi^{(\mathrm{ETA})}=V^{(\mathrm{ETA})\dag}\Sigma^{(\mathrm{ETA})}V^{(\mathrm{ETA})},
\end{equation}
where $\Sigma^{(\mathrm{ETA})}$ contains the eigenvalues of the $|\tilde{\Psi}_j\rangle$.
One has a freedom in choosing the unitary $V^{(\mathrm{ETA})}$ (as one can always re-label indices on the $|\Psi_j\rangle$); for our purposes, we want to choose $V^{(\mathrm{ETA})}$ to maximise the overlap $|\langle\Psi_j|\tilde{\Psi}_j\rangle|^2=|V_{j,j}|^2$.
Let us now for simplicity focus on the second-order derivative $\partial^2 E/\partial\lambda_1\partial\lambda_2$, and fix $\hat{P}=\partial\ham/\partial\lambda_1$, $\hat{Q}=\partial\ham/\partial\lambda_2$.
One may rewrite the term $\sum_j\mathcal{A}_1(j)f(E_j)$ (Eq.~6 of the main text) as
\begin{equation}
\sum_{j,l}p^*_jF_{j,l}q_l,
\end{equation}
where $F(j,l)=\delta_{j,l}f(E_j)$, $p_j=\langle\Psi_j|\hat{P}|\Psi_0\rangle$, and $q_l=\langle\Psi_j|\hat{Q}|\Psi_0\rangle$.
In the response method, this is replaced by
\begin{equation}
\sum_{j,l}\tilde{p}^*_j\tilde{F}_{j,l}\tilde{q}_l=\sum_{j,l,m,n,o,p}p^*_j\Pi_{j,m}V_{m,n}\tilde{F}_{n,o}V^{\dag}_{o,p}\Pi_{p,l}q_l.
\end{equation}
The error in the approximation can then be split into:
\begin{enumerate}
\item The error in the projection $\Pi$: in general $\|\Pi\vec{p}\|\leq \|\vec{p}\|$, with equality when $\Pi\vec{p}=\vec{p}$ (in which case this error source does not exist).
\item The error in the unitary rotation $V$: ideally $V=I$, or at least $V\Pi\vec{p}=\Pi\vec{p}$ (in which case this error source does not exist).
\item The error in the energy function $\tilde{F}$: ideally $\tilde{F}=F$, or at least $\tilde{F}V\Pi\vec{p}=FV\Pi\vec{p}$ (in which case this error source does not exist).
\end{enumerate}
Generic bounds on these three error sources or measurements thereof are difficult to come by.
The first may be measured directly when $\hat{P}$ or $\hat{Q}$ is a unitary operator, as then $\|\vec{p}\|=1$ and $\|\Pi\vec{p}\|$ may be measured.
Moreover, for the QSE approximation, if $\hat{P}=\sum_i p_i\hat{E}_i$ (where $\hat{E}_i$ are the excitation operators), then $\hat{P}|\Psi_0\rangle$ lies within $\Hh^{(\mathrm{ETA})}$ and this error source is cancelled.
The second may be estimated partially by determining the uncertainty in the energies of the approximate eigenstates
\begin{equation}
\sigma^2_{|\tilde{\Psi}_j\rangle,\ham}=\langle\tilde{\Psi}_j|\ham^2|\tilde{\Psi}_j\rangle-\langle\tilde{\Psi}_j|\ham|\tilde{\Psi}_j\rangle^2.
\end{equation}
However, this underestimates the uncertainty from eigenstates that are of similar energy (but may be not connected by $\hat{P}\hat{Q}$).
One may further estimate this by measuring $\langle\Psi_0|\hat{P}\hat{Q}|\Psi_0\rangle$ and comparing to the approximation $\langle\Psi_0|\hat{P}^{(\mathrm{ETA})}\hat{Q}^{(\mathrm{ETA})}|\Psi_0\rangle$ (although this in turn may miss favourable cancellations of different amplitudes).
However, if one estimates $|\tilde{\Psi}_j\rangle$ via direct approximation methods, this source of error should be similarly negligible.
In this case the final source of error is also negligible (if one can both approximate $|\tilde{\Psi}_j\rangle=|\Psi_j\rangle$ and estimate the energy $E_j$).
For the QSE, the final source of error is also bounded above by $\sigma^2_{|\tilde{\Psi}_j\rangle,\ham}$, which is in turn bounded by the maximum energy of the excitation operators $\|[\ham,E_j]\|$.
Making these bounds more precise and tighter for various excited state methods is a clear direction for future research.

\section{Amplitude estimation error for single-round QPE}
In this section, we extend the analysis of~\cite{obrien2019quantum}(App.B) to bound the error in estimating an amplitude $A_j=|a_j|^2$ via single-ancilla QPE.
As mentioned in the main text, this is a loose bound, and we expect single-ancilla QPE to achieve similar scaling to QPE performed with multiple ancillas.
To calculate the bound, let us consider the situation of attempting to estimate a single phase $\phi$ and a corresponding (complex-valued) amplitude $a$ from terms of the function
\begin{equation}
g(k)=A e^{ik\phi},
\end{equation}
for $k=0,\ldots,k_{\max}$.
(Note that $k_{\max}$ corresponds to $K$ in~\cite{obrien2019quantum})
We assume the real and imaginary part of $g(k)$ are estimated independently as $g_k=g_k^0+ig_k^1$ from $\nmeas$ single-shot measurements.
Following the procedure of~\cite{obrien2019quantum}, one first estimates the phase $\phi$ via least squares as
\begin{equation}
\tilde{\phi}=\frac{\sum_{k=0}^{k_{\max}}g^*_kg_{k+1}}{\sum_kg^*_kg_k}.
\end{equation}
From here, one may construct the vector $f_k=e^{ik\tilde{\phi}}=f_k^0+if_k^1$ and estimate the amplitude $A$ via least squares as
\begin{equation}
\tilde{A}=\frac{1}{L}\frac{\sum_{k=1}^Lf_k^*g_k}{\sum_{k=1}^Lf_k^*f_k}=\frac{1}{L}\sum_{k=1}^Lf_k^*g_k=A_0+iA_1.
\end{equation}
(We may evaluate the denominator immediately as, regardless of any error in $\phi$, $f_k$ will satisfy $|f_k|^2=1$.)
We do not necessarily wish to include all $k\in\{1,\ldots,k_{\max}\}$ in this estimation, as the error in $f_k$ at large $k$ is smaller than the error at small $k$.
To make things worse, errors in $f_k^a$ are correlated.
It was derived in~\cite{obrien2019quantum} that this correlation depends only on $g_{k_{\max}}$; adjusting for the fact that $|A|\leq 1$, we obtain
\begin{align*}
\frac{\partial f_k^a}{\partial g_{k'}^{a'}}\propto\delta_{k',k_{\max}}\frac{k}{k_{\max}|A|}
\end{align*}
Propagating variances then obtains
\begin{align*}
\mathrm{Var}[\tilde{A}_0]&=\frac{1}{L^2}\sum_{k=1}^L\sum_{a=0,1}|f_k^a|^2\mathrm{Var}[g_k^a]\\&+\frac{1}{L^2}\sum_a\left[\sum_{k=1}^L\frac{\partial f_k^a}{\partial g_{k_{\max}}^a}g_k^a\right]^2\mathrm{Var}[g_{k_{\max}}^a]\\
&\approx\frac{1}{L^2}\sum_{k=1}^L\frac{1}{N}+\frac{1}{L^2}\left[\frac{L^2}{k_{\max}|A|^2}|A|^2\right]^2\frac{1}{N}\\
&\approx\frac{L^2}{k_{\max}^2N}+\frac{1}{LN},
\end{align*}
and a similar scaling for $\mathrm{Var}[\tilde{A}_1]$
We then choose $L$ to minimize the variance
\begin{equation}
\frac{\partial\mathrm{Var}[\tilde{A}]}{\partial L}=2\frac{L}{k_{\max}^2N}-\frac{1}{L^2N}=0\rightarrow L\propto k_{\max}^{\frac{2}{3}},
\end{equation}
resulting in a final variance scaling as
\begin{equation}
\mathrm{Var}[\tilde{A}]\propto \frac{1}{k_{\max}^{\frac{2}{3}}N}.
\end{equation}

\end{document}